\documentclass[fleqn,usenatbib]{mnras}

\usepackage{newtxtext,newtxmath}

\usepackage[T1]{fontenc}

\DeclareRobustCommand{\VAN}[3]{#2}
\let\VANthebibliography\thebibliography
\def\thebibliography{\DeclareRobustCommand{\VAN}[3]{##3}\VANthebibliography}

\usepackage{graphicx}	
\usepackage{amsmath}	

\title[Detection of radio emission of 20 pulsars]{The second decametre pulsar census at UTR-2 radio telescope}

\author[I. P. Kravtsov et al.]{
Ihor P. Kravtsov,$^{1}$\thanks{E-mail: i.p.kravtsov@gmail.com (KTS)}
Vyacheslav V. Zakharenko,$^{1,2}$
Oleg M. Ulyanov,$^{1}$
Alisa I. Shevtsova,$^{1}$
\newauthor Serge M. Yerin,$^{1,2}$ Oleksandr. O. Konovalenko$^{1}$
\\
$^{1}$Institute of Radio Astronomy of NAS of Ukraine, 4 Mystetstv st., 61002, Kharkiv, Ukraine\\
$^{2}$V. N. Karazin Kharkiv National University, 4 Svobody Sq., 61022, Kharkiv, Ukraine
}

\date{Accepted XXX. Received YYY; in original form ZZZ}

\pubyear{2021}

\begin{document}
\label{firstpage}
\pagerange{\pageref{firstpage}--\pageref{lastpage}}
\maketitle

\begin{abstract}
Our paper presents the results of the second census of pulsars in decametre wave range at UTR-2 radio telescope. Over the past ten years, the number of discovered nearby pulsars in the world has doubled, which has made it urgent to search for a low-frequency radio emission from newly discovered sources. To increase this census sensitivity, the integration time was doubled compared with the first census of 2010-2013. As a result, the decametre radio emission of 20 pulsars was detected, their flux densities and the shape of pulses were obtained. The dispersion measure for 10 pulsars and the rotation period for 8 pulsars were refined. For several pulsars the scattering time constant and FWHM were estimated in decametre wave range. Upper limits of flux densities of 102 not yet detected pulsars were also estimated.

\end{abstract}




\section{Introduction}

Pulsars were discovered more than 50 years ago \citep[][] {Hewish1968} however, the interest in studying them is constantly growing due to their importance for studies of interstellar medium, stellar evolution, high-energy physics and possible gravitational waves sensing.

Due to the development of radio astronomical observation techniques and data processing tools (as well as ongoing efforts in search for new sources), the number of discovered pulsars is constantly increasing. The maximum flux of these sources’ radiation lays in the meter wave range, that means the new searches of nearby sources in the frequency range below 1\,GHz are quite promising. A lot of new instruments have been introduced in this range recently. Among them UTMOST (upgraded telescope Molongo Observatory Synthesis Telescope) \citep{Bailes2017}, LWA (Long Wavelength Array) \citep{Stovall2015}, NenuFAR (New Extensionin Nançay Upgrading LOFAR) \citep[][]{Zarka2015} can be named. The most anticipated in the whole radio range is the introduction of SKA (Square Kilometer Array)  \citep[][]{Keane2014}.

For the extremely low frequencies (10$-$30\,MHz), the strong increase in the scattering time of pulsar pulses with a frequency decreasing ($\propto\nu^{-4.4}$) makes it impossible to study the pulsed radiation of distant pulsars. These studies are further complicated by factors such as: (i) larger dispersion delay comparing to its values at high frequencies ($ \propto f^{-2}$), (ii) increase in the background brightness temperature of the Galaxy ($ \propto f^{-2.55}$) \citep{Lawson1987}, (iii) growing power and number of terrestrial radio frequency interference sources. 

However, for nearby pulsars with dispersion measures (DM) up to $30$\,pc\,cm$^{-3}$, which correspond to a distance of 1-2\,kpc (depending on the pulsar galactic coordinates), the number of pulsars detected before 2010 in the decametre wave range (about a dozen) seemed very small. Deep upgrade of the UTR-2 radio telescope and its receiving system \citep{Konovalenko2016, Zakharenko2016}. allowed to increase this number. It turned out that for effective pulsar detection in the decametre wave range, the DM values obtained from high frequency data in some cases are not accurate enough.  Refinement of the DM values during data processing, maximal use of the bandwidth and radio frequency interference (RFI) mitigation in the first decametre pulsar census (2010-2013) allowed authors to detect 40 of 74 known at that time pulsars with parameters: rotation period ($P$) longer than $0.1$\,s, DM less than $30$\,pc\,cm$^{-3}$, declinations ($\delta$ or dec) above\,$-10^{\circ}$ \citep{Zakharenko2013}.

The first decametre survey of pulsars and sporadic radiation sources of the Northern sky \citep{Vasylieva2013, Vasylieva2015,  Kravtsov2016a, Kravtsov2016b, Kravtsov2016c, Zakharenko2018} revealed a large number of bursts originating from space (non-repeating pulses that look like pulsar emission). Some of the bursts could be single (anomalously intense) pulses \citep{Ulyanov2006, Ul'yanov2012} of hitherto unknown pulsars, but to compare and reliably identify these pulses with known sources, it is necessary to determine their radiation parameters with sufficient accuracy. For example, the DM value must be measured with an accuracy of $0.01$\,pc\,cm$^{-3}$ or higher.

In the last decade the number of the new pulsars with parameters suitable for observations at the UTR-2 has doubled, especially due to such surveys as the LOTAAS (at LOFAR) \citep{Sanidas2019} and the GBT pulsar survey (at the Green Bank Telescope) \citep{Kawash2018}. The new discoveries demanded a new census of pulsars in the decametre wave range both to expand the list of studied sources and to potentially identify the discovered transient signals with the individual pulses of new pulsars.

\section{Instrumentation and observations}
\label{sec:Instrumentation and observations} 

For pulsar observations at the UTR-2 radio telescope \citep{Konovalenko2016} we used the sum of “North – South” and “East –West” antennas’ signals mode. This mode provides the highest sensitivity, while the confusion effect during the pulsar search is not dangerous. In contrast to the first decametre pulsar census \citep{Zakharenko2013}, when the question of the possibility of detecting the radiation of a large number of pulsars at UTR-2 was actually relevant, in this work we pay more attention to the sensitivity increasing by means of increasing the integration time from $1.5$\,hours in the previous census to $3$\,hours in the current one. As in \citet{Zakharenko2013}, a range of dispersion measures is 0÷30\,pc\,cm$^{-3}$ with a step of $0.002$\,pc\,cm$^{-3}$ in 16.5$-$33.0\,MHz frequency range (4096 channels with $\approx$ $4$\,kHz width) was chosen. Since the scattering time constant in the decametre wave range for majority of known pulsars (even the nearest ones) exceeds $10$\,ms, we use a time resolution of $8$\,ms, a post-detector method of dedispersion and an advanced \citep[in comparison with][]{Zakharenko2013} RFI mitigation procedure \citep{Vasylieva2013, Vasylieva2015}.

As of the beginning of 2020, 163 pulsars which meet the abovementioned criteria (DM\,<\,$30$\,pc\,cm$^{-3}$, $P$\,$>$\,$0.1$\,s, $\delta$$>$\,-10$^{\circ}$) were known from the primary sources \citep[see][and references there]{Kravtsov2020} and from the ATNF pulsar catalogue \citep {Manchester2005}. Forty of these pulsars were detected during the first census, another one (PSR J0242+62) was detected later \citep{Vasylieva2014}. Therefore, in this work we observed 122 pulsars, which have not been detected at decametre waves yet. Note that some characteristics of a number of recently discovered pulsars has not been determined (flux density, scattering time constant, etc.) or has not been measured accurately enough. Among these parameters the main one is the pulsar period therefore it was taken as a tunable parameter for some pulsars during their search.

\begin{figure}
	\includegraphics[width=\columnwidth]{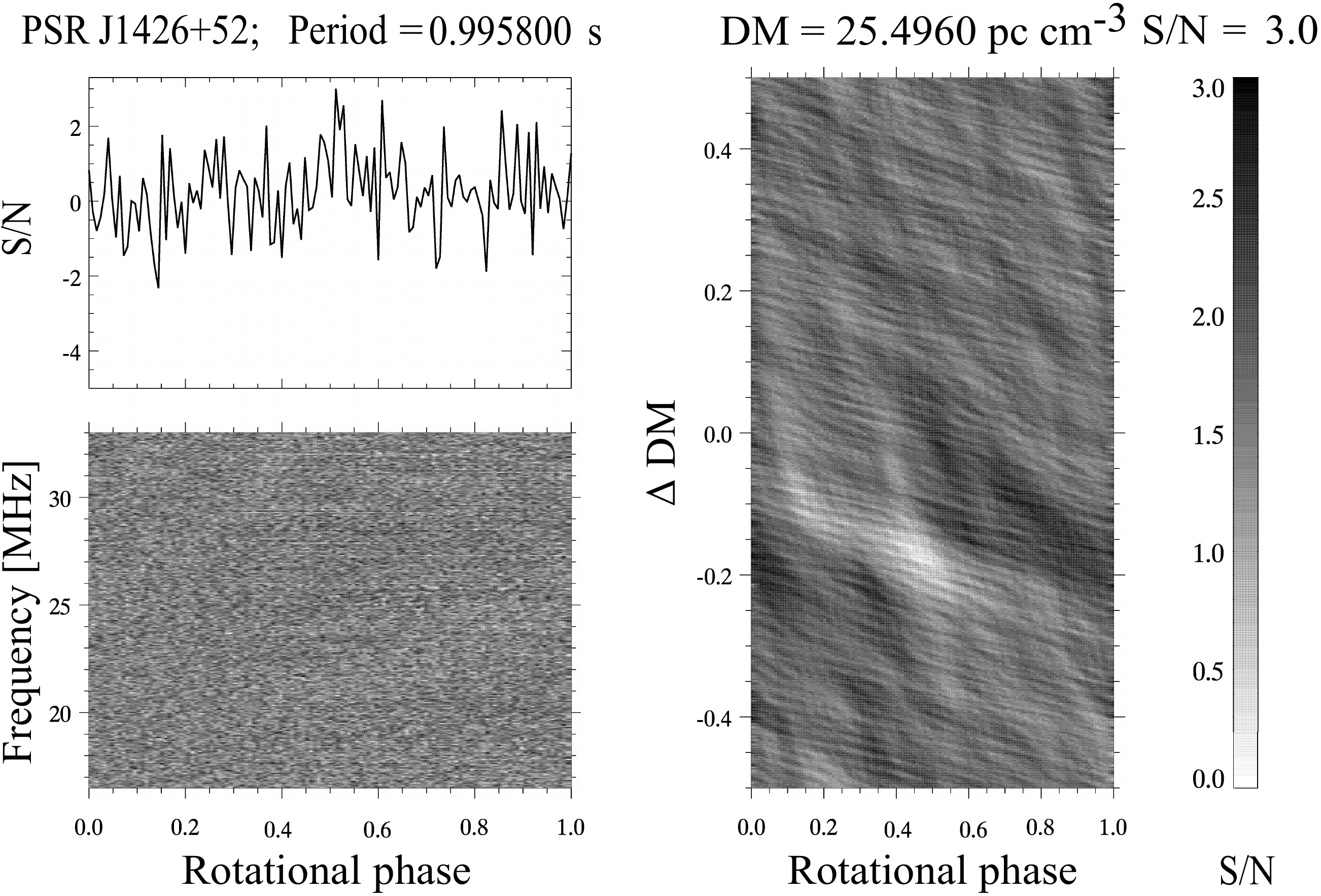}
    \caption{Results of PSR\,J1426+52 data processing with the rough period value $P$\,=\,0.9958\,s (hereinafter intensity is normalized to the off-pulse noise). The upper left panel shows an average pulse profile, the lower left one – its average spectrum. The right panel shows "dispersion measure\,–\,rotational phase" plot, as in \citet{Zakharenko2013}, which helps to refine the DM value.}
    \label{fig:PSRJ1426+52_1.pdf}
\end{figure}

\begin{figure}
	\includegraphics[width=\columnwidth]{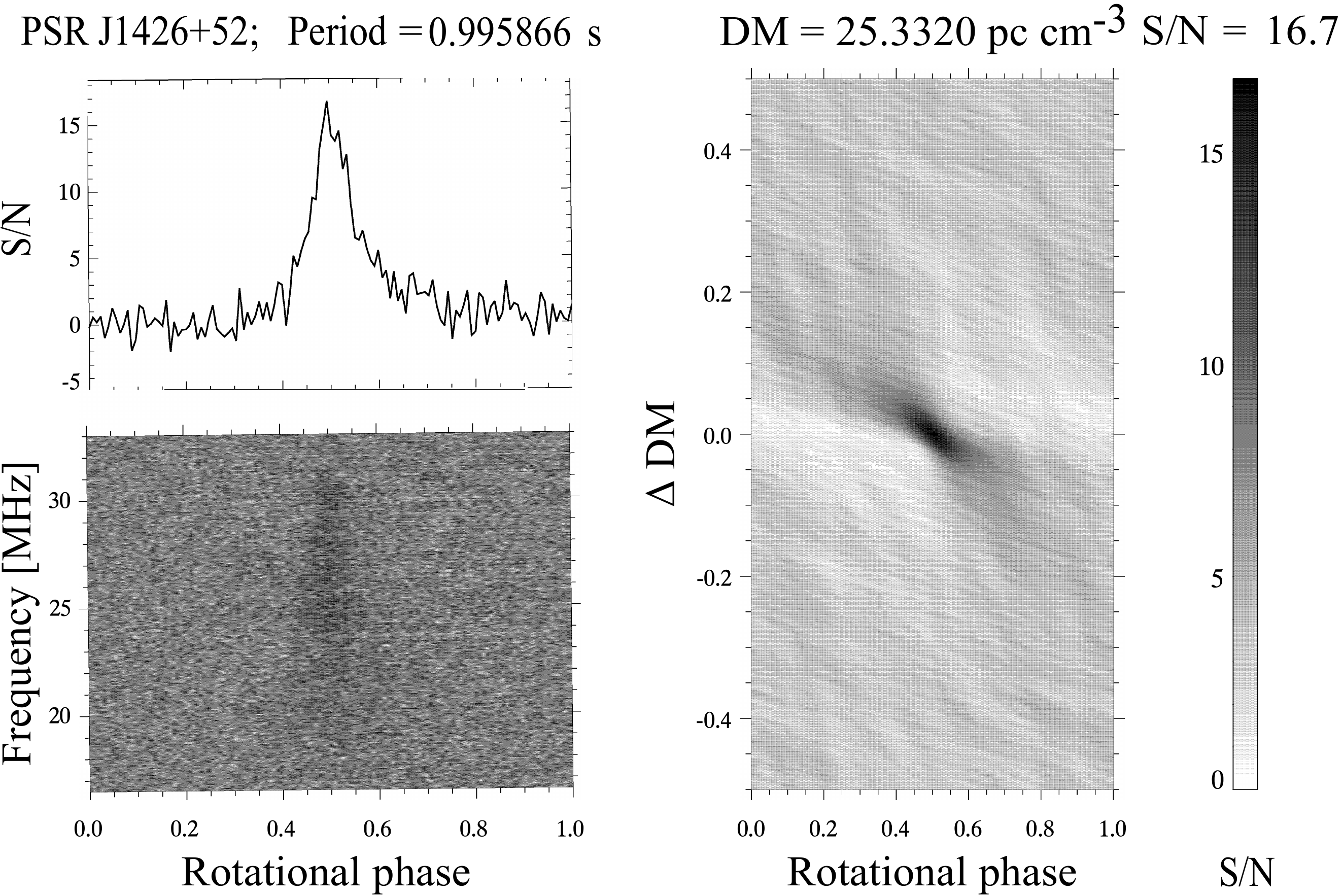}
    \caption{Results of detection of PSR\,J1426+52 with the refined period value $P$\,=\,0.995866\,s.}
    \label{fig:PSRJ1426+52_2.pdf}
\end{figure}

As an example of observational data processing of a pulsar with poor rotational period accuracy we can show the PSR\,J1426+52 before our refinement (Fig.~\ref{fig:PSRJ1426+52_1.pdf}) and successful detection after the period refinement (Fig.~\ref{fig:PSRJ1426+52_2.pdf}). The rotation period of PSR\,J1426+52 reported in \citet{Sanidas2019} has only four significant digits ($0.9958$\,s), so its accuracy is rather low, same as for dozens of other recently discovered pulsars. Using of this rough rotation period value gave the result shown in Fig.~\ref{fig:PSRJ1426+52_1.pdf}.

One should note that the "DM\,–\,rotational phase" plot has light inclined oval centered at 0.48 on the x-axis and $0.18$\,pc\,cm$^{-3}$\,–\,on y-axis. With clarifying the pulsar period with coarse (10$^{-5}$\,s) and fine (10$^{-6}$\,s) steps to speed up calculations, we detected PSR\,J1426+52 with the maximal signal-to-noise ratio (S/N) of about 17 and the narrowest average pulse profile with $P$\,=\,$0.995866$\,s \citep{Kravtsov2020}.

\begin{figure}
	\includegraphics[width=\columnwidth]{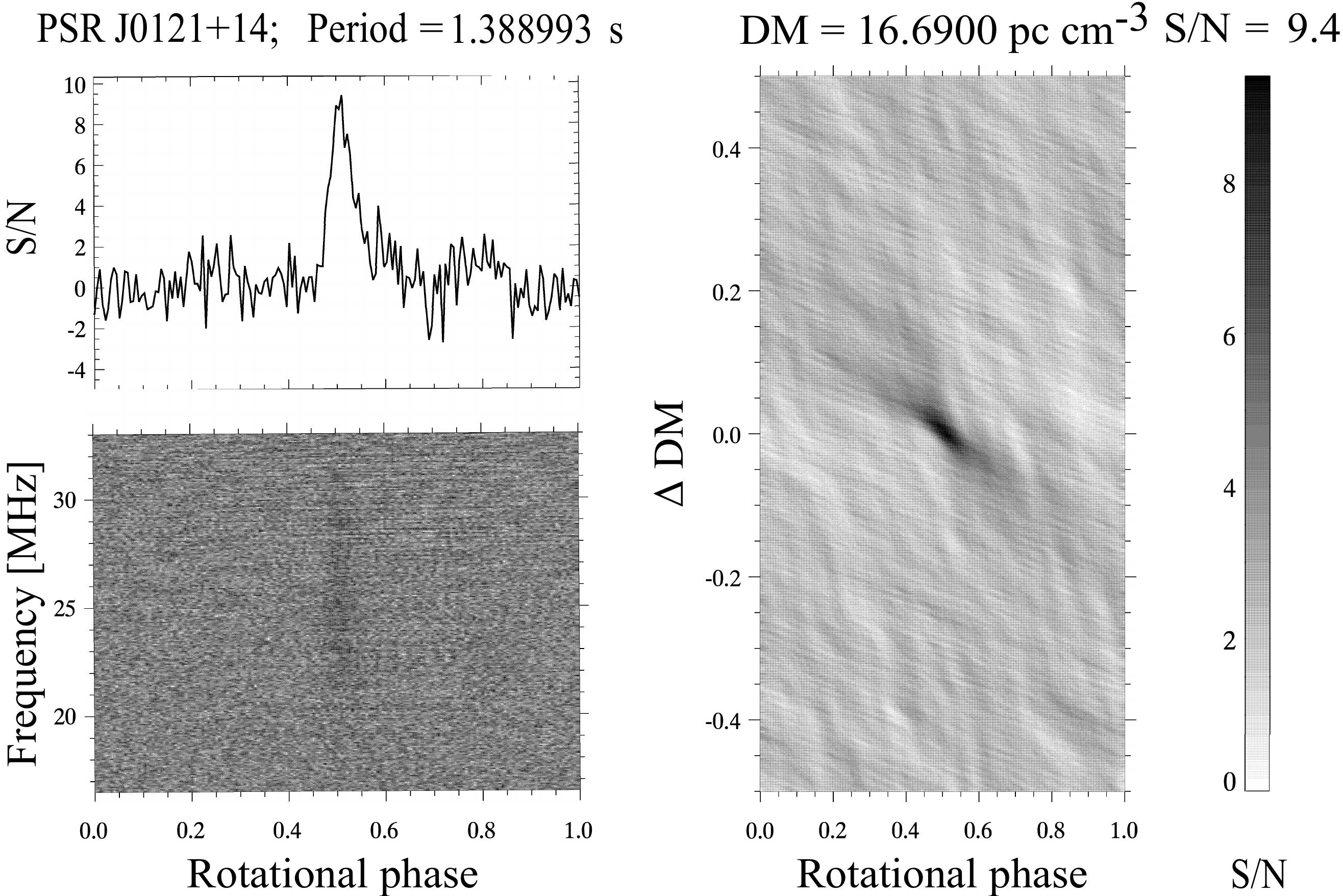}
    \caption{Results of PSR\,J0121+14 detection on February 10, 2020 with the refined rotation period $P$\,=\,1.388993\,s.}
    \label{fig:PSRJ0121+14.pdf}
\end{figure}

Another example of rotational period refinement is PSR\,J0121+14. It was detected in the daytime (its culmination on February 10, 2020 was around 15:30 local time). The measured DM value is $17.69$\,pc\,cm$^{-3}$ (hereinafter the error in DM determination obtained when the S/N falls to 0.95 from the maximal value) and does not differ from that obtained in \citet{Sanidas2019} (DM\,=\,$17.77$\,pc\,cm$^{-3}$). We have also refined the pulsar period (our value $P$\,=\,$1.388993$\,s, see Fig.~\ref{fig:PSRJ0121+14.pdf}) which was indicated with lower accuracy in \citet{Sanidas2019} a $1.3890$\,s. Some time later, in \citet{Michilli2020} rotational period of this source was measured with high accuracy ($P$\,=\,$1.38899395038$\,s). Our rotational period value completely coincides with the high-precision measurements from \citet{Michilli2020} (taking into account our restriction described at the beginning of the Section \ref{sec:Results and discussion}), which confirms the reliability of our refinement of the rotational period using the criterion for maximizing the S/N and obtaining the narrowest possible pulse profile of the pulsar. Since PSR\,J0121+14 is quite a powerful source, we also succeeded to estimate its scattering time constant, which is $\tau_{sc}$\,=\,51$\pm$$16$\,ms. Scattering time constant was determined following \citet{Zakharenko2013} by fitting the intensity decrease versus time in the pulse tail by a law $\propto e^{-t/\tau_{sc}}$, assuming a rectangular shape for the pulse.

In order to optimize the observation time at the UTR-2 radio telescope, the observation sessions each time included a set of pulsars with culmination times spaced by about $3$\,hours from one another. Such pulsar selection provides maximal use of the UTR-2 effective area, because each source was recorded close to its culmination. However, observations of the majority of pulsars were carried out in the morning, afternoon and evening, i.e. in suboptimal interference conditions. Note that in the first census about 60\% of observations were made during the daytime, which in the presence of a strong RFI in the decametre wave range could be an obstacle to the detection of a significant number of pulsars. In this work, the pulsars were observed in a period from February 2020 to February 2021 (February 3-17, May 25-June 1, September 7-14, November 16-23 and December 21-26 – in 2020 and February 8-15, 2021).

\section{Results and discussion}
\label{sec:Results and discussion}

As a result of the work, the decametre radio emission of 20 pulsars has been detected. For 16 pulsars, the values of their DMs have been refined. In 8 cases, the period was refined up to the sixth significant digit (based on the ratio of time resolution and duration of observations: $0.008$\,s\,/\,$10800$\,s~=~7.4·$10^{-7}$). For some pulsars, the scattering time constant at the central frequency of the range ($24$\,MHz) has been determined. We provide the data for each pulsar below.

\subsection{Average pulse profile and pulsar parameters}
\label{sec:Average pulse profile and pulsar parameters} 

{\itshape {PSR\,J0121+14}}. We have already shown this pulsar in Section \ref{sec:Instrumentation and observations} (see Fig.~\ref{fig:PSRJ0121+14.pdf}) as an example of period refinement, we found its $P$\,=\,$1.388993$\,s and DM\,=\,17.69$\pm$$0.02$\,pc\,cm$^{-3}$.

\begin{figure}
	\includegraphics[width=\columnwidth]{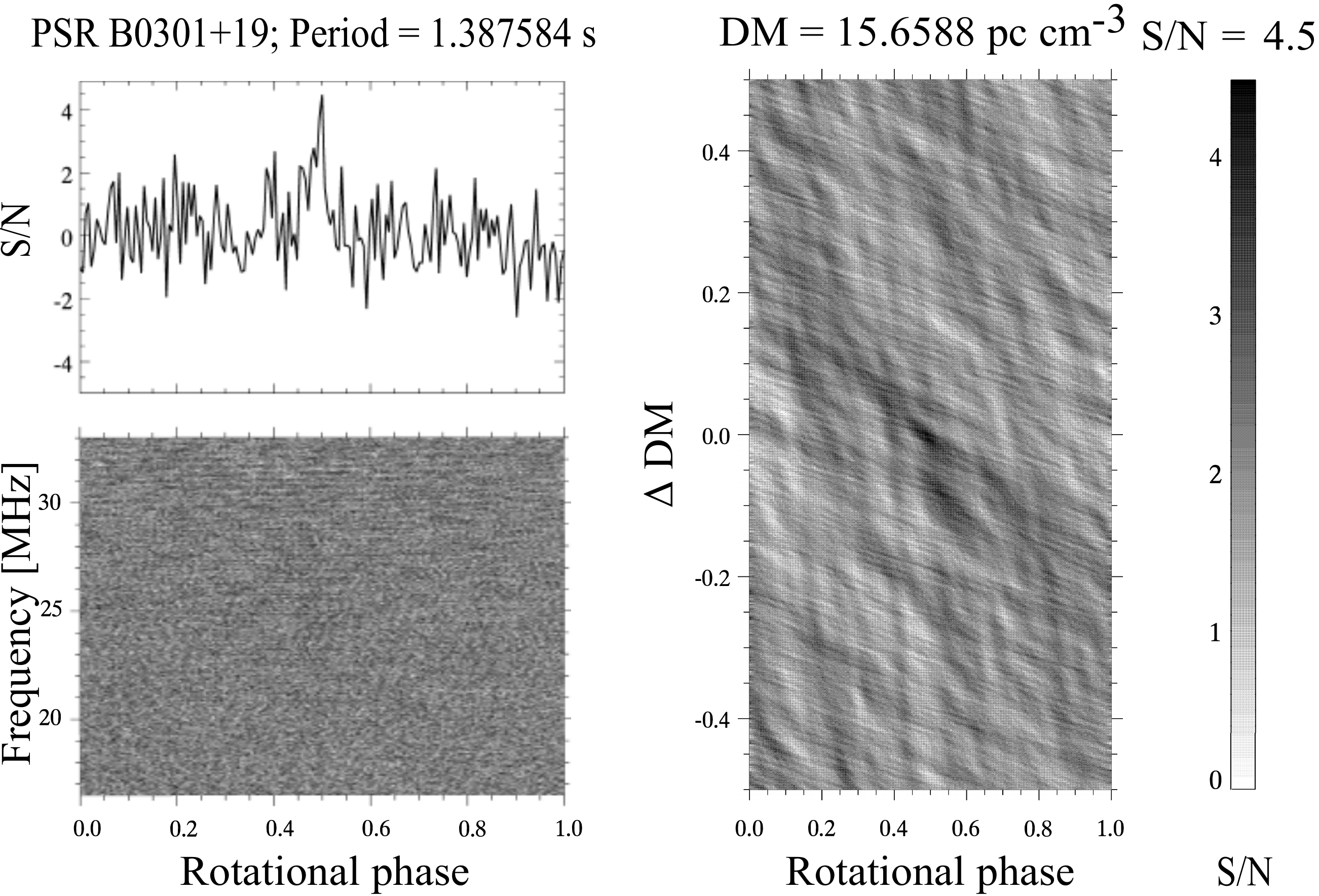}
    \caption{Results of PSR\,B0301+19 detection of on February 5, 2020.}
    \label{fig:PSRB0301+19.pdf}
\end{figure}

{\itshape {PSR\,B0301+19}} was detected with a low S/N of about 5 (Fig.~\ref{fig:PSRB0301+19.pdf}). It correlates well with the results of\;\citet{Zakharenko2013}, where 1.5-hour observations did not result in this pulsar radio emission detection above the level ($4\sigma$). This result allows us to compare the sensitivity of the two decametre pulsar censuses and to estimate the upper detection limit \citep[see paragraph 3.2 in ][]{Zakharenko2013}. We measured the DM to be 15.659$\pm$$0.020$\,pc\,cm$^{-3}$, taking into account the error, it does not differ from the value 15.65677$\pm$$0.00035$\,pc\,cm$^{-3}$) specified in \citet{Bilous2016}.

\begin{figure}
	\includegraphics[width=\columnwidth]{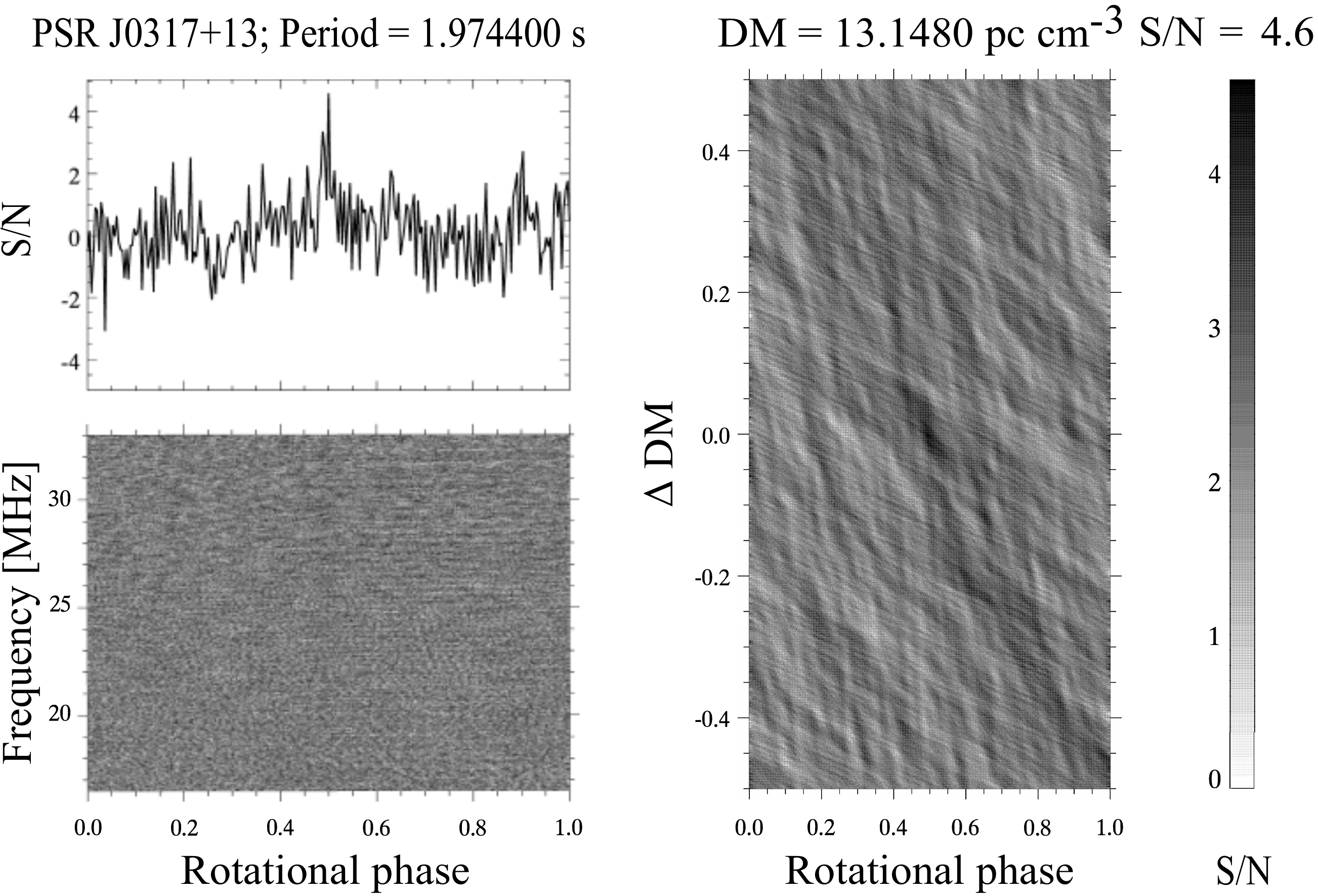}
    \caption{Results of PSR\,J0317+13 detection on February 16, 2020 with the refined rotation period $P$\,=\,1.97440\,s.}
    \label{fig:PSRJ0317+13.pdf}
\end{figure}

{\itshape {PSR\,J0317+13}}. When we processed the data of PSR\,J0317+13 with parameters given in \citet{Sanidas2019}, namely the period $P$\,=\,$1.9743$\,s and the DM\,=\,12.90$\pm$$0.04$\,pc\,cm$^{-3}$, we were not able to detect this pulsar. However, the period value variation with a step of $0.00001$\,s led to a weak detection of an average pulse profile at $P$\,=\,1.97440\,s (Fig.~\ref{fig:PSRJ0317+13.pdf}) and allowed us to refine the DM to 13.148$\pm$$0.020$\,pc\,cm$^{-3}$.

\begin{figure}
	\includegraphics[width=\columnwidth]{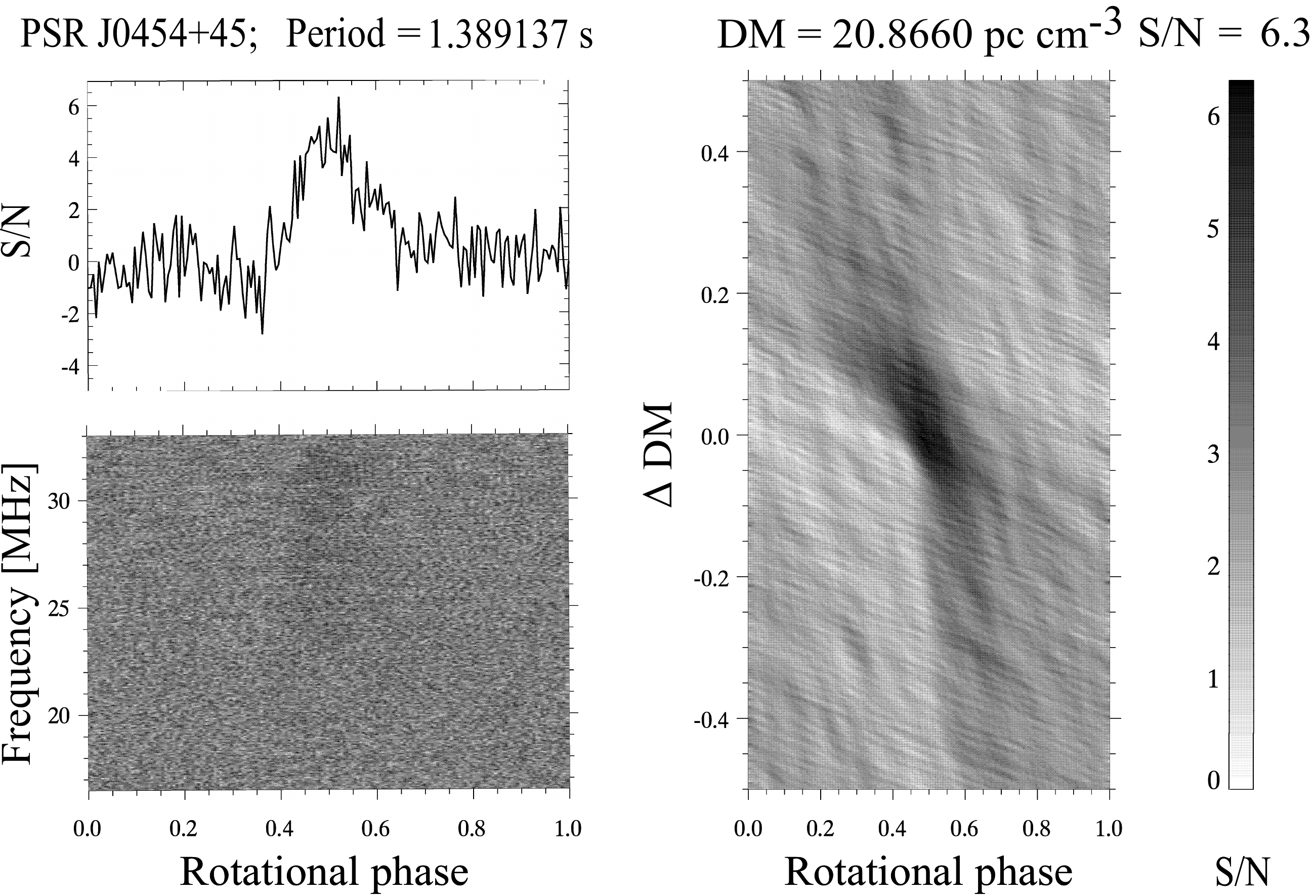}
    \caption{Detection of PSR\,J0454+45 on February 8, 2021.}
    \label{fig:PSRJ0454+45.pdf}
\end{figure}

{\itshape {PSR\,J0454+45}}. We used $P$\,=\,1.3891369360\,s and DM\,=\,20.834$\pm$$0.002$\,pc\,cm$^{-3}$ from \citet{Tan2020} to process the data and obtained the average profile shown in Fig.~\ref{fig:PSRJ0454+45.pdf}. We estimated the DM\,=\,20.854$\pm$$0.020$\,pc\,cm$^{-3}$ which is very close to the values from the paper.

\begin{figure}
	\includegraphics[width=\columnwidth]{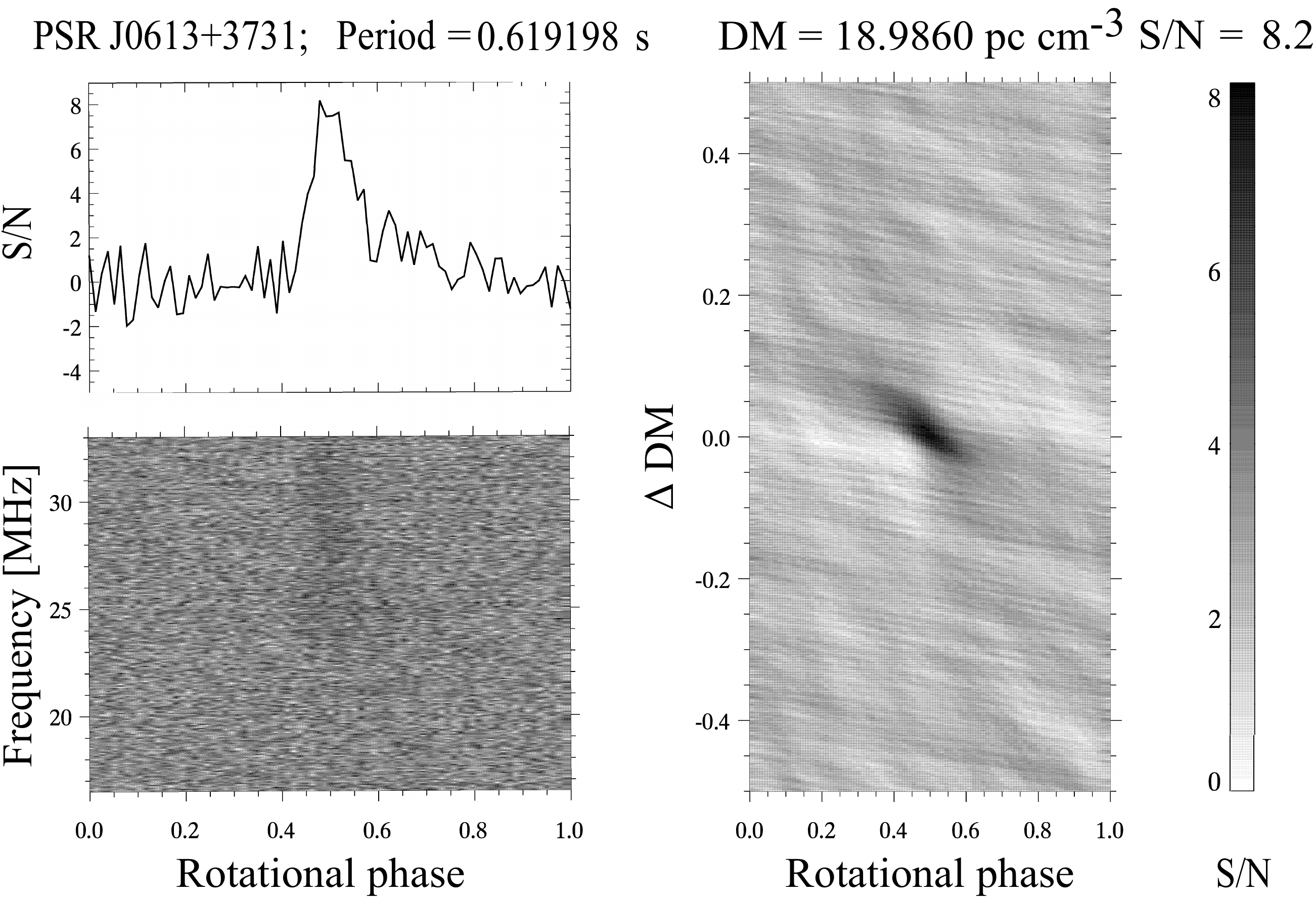}
    \caption{Detection of PSR\,J0613+3731 on November 19, 2020.}
    \label{fig:PSRJ0613+3731.pdf}
\end{figure}

{\itshape {PSR\,J0613+3731}} was observed three times during the year: in February, September and November 2020. The third session resulted in the most reliable detection (S/N\,$>$\,8) due to the most suitable for observations season of the year and culmination local time at about 1h\,50m. Our average pulse profile (Fig.~\ref{fig:PSRJ0613+3731.pdf}) allowed us to refine the DM\,=\,18.986$\pm$$0.020$\,pc\,cm$^{-3}$ which almost equals to the one specified in \citet{Tan2020} (18.990$\pm$$0.012$\,pc\,cm$^{-3}$) taking errors into account. We also estimated the pulse's scattering time constant, which is $\tau_{sc}$\,=\,30$\pm$$10$\,ms. 

\begin{figure}
	\includegraphics[width=\columnwidth]{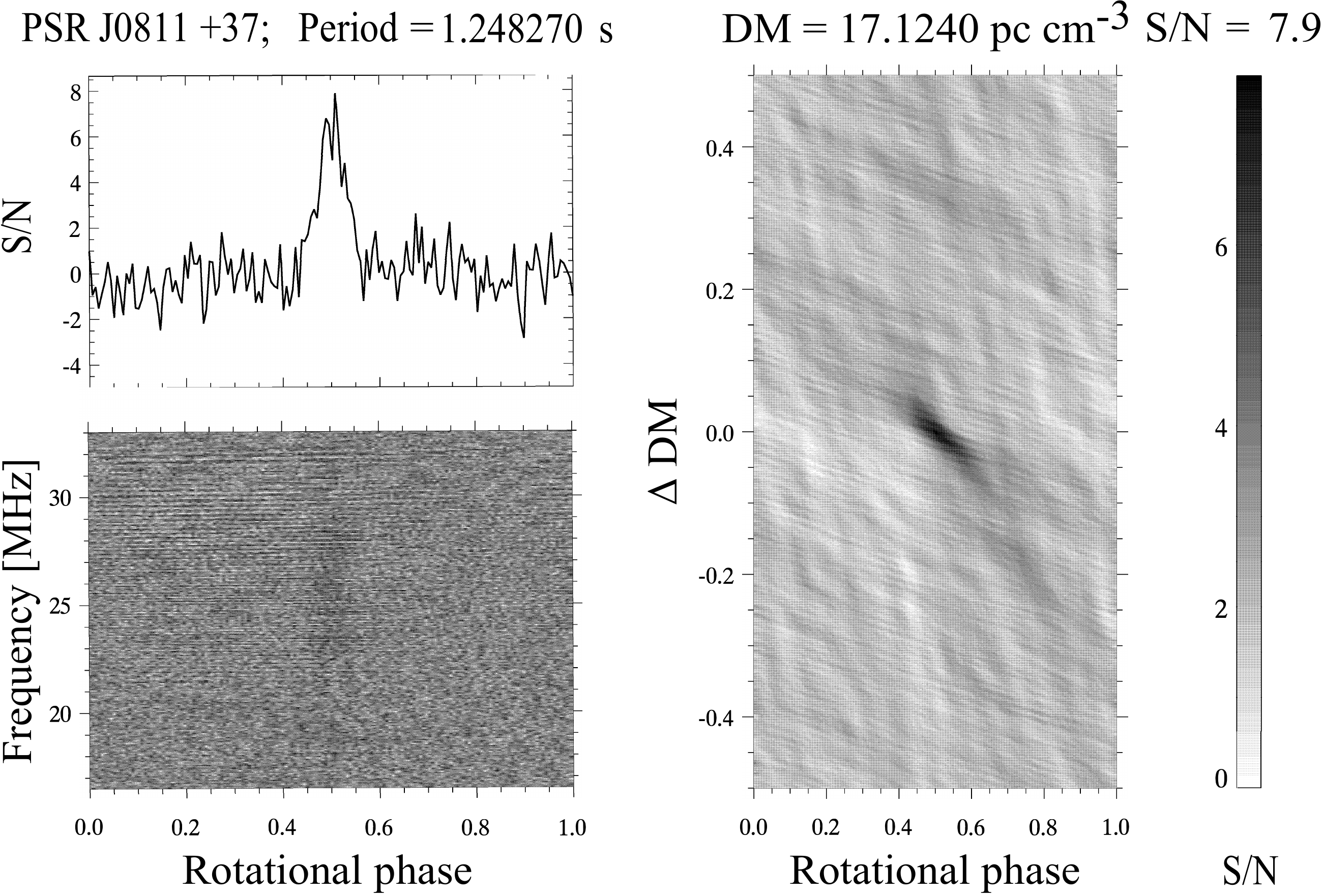}
    \caption{Detection of PSR\,J0811+37 on February 9, 2020 with the refined rotation period $P$\,=\,$1.24827$\,s.}
    \label{fig:PSRJ0811+37.pdf}
\end{figure}

{\itshape {PSR\,J0811+37}}. We used initial parameters DM\,=\,16.95$\pm$$0.11$\,pc\,cm$^{-3}$ and $P$\,=\,$1.2483$\,s from \citet{Sanidas2019}, but obtained the best result – a narrow average pulse profile with S/N\,>\,7 shown in Fig.~\ref{fig:PSRJ0811+37.pdf} - with refined parameters of $P$\,=\,$1.24827$\,s (close to the initial one) and DM\,=\,17.124$\pm$$0.020$\,pc\,cm$^{-3}$, that is slightly bigger than the initial value. Our estimation of its scattering time constant $\tau_{sc}$\,=\,33$\pm$$10$\,ms.

\begin{figure}
	\includegraphics[width=\columnwidth]{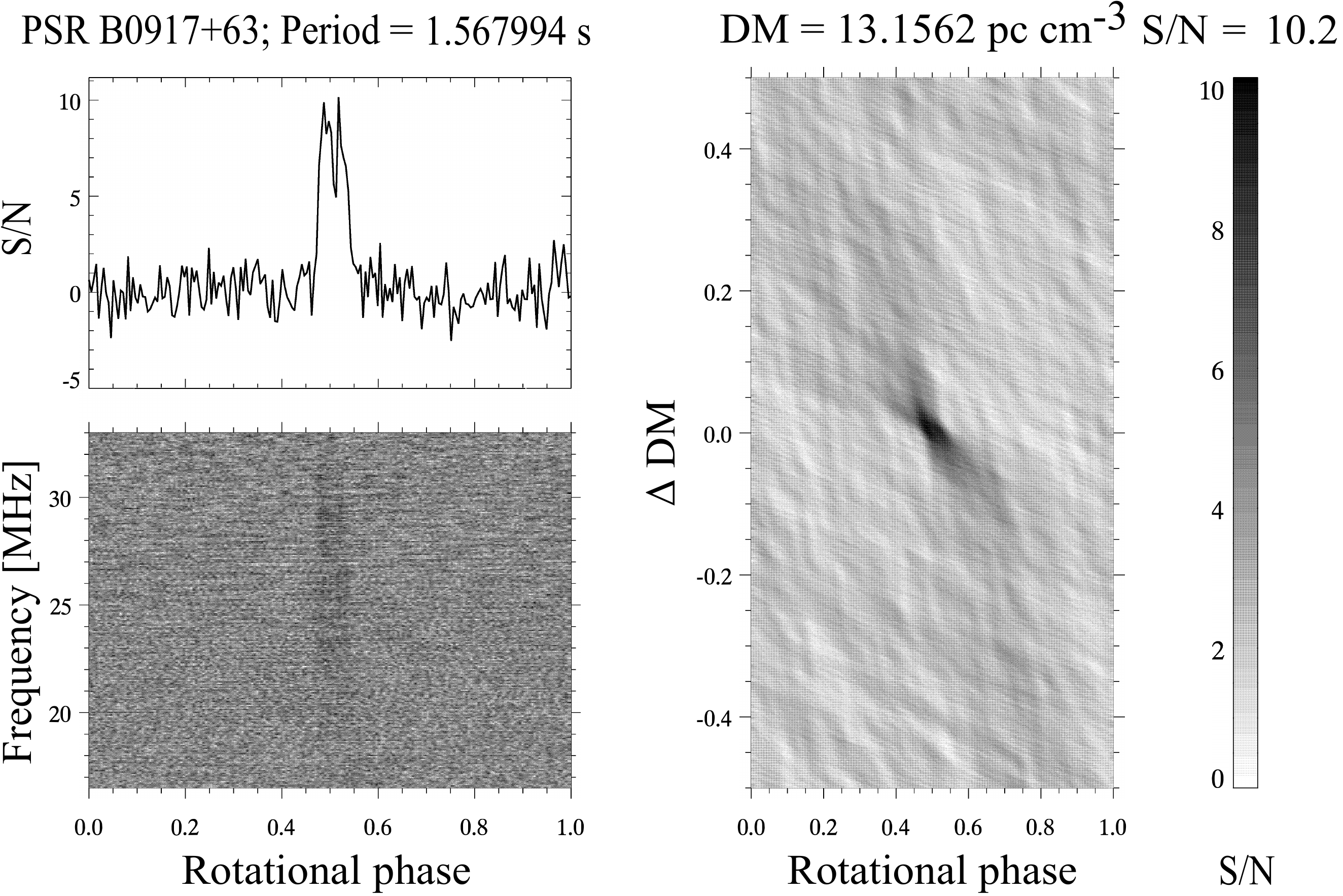}
    \caption{Detection of PSR\,B0917+63 on May 28, 2020.}
    \label{fig:PSRB0917+63.pdf}
\end{figure}

{\itshape {PSR\,B0917+63}}. We used parameters DM\,=\,13.15423$\pm$$0.00018$\,pc\,cm$^{-3}$, and the period $P$\,=\,1.567994018\,s from \citet{Bilous2016} and \citet{Hobbs2004} for data processing, the prominent average pulse profile (Fig.~\ref{fig:PSRB0917+63.pdf}) allowed us to obtain the DM\,=\,13.156$\pm$$0.010$\,pc\,cm$^{-3}$ and estimate the scattering time constant $\tau_{sc}$\,=\,15$\pm$$8$\,ms.

\begin{figure}
	\includegraphics[width=\columnwidth]{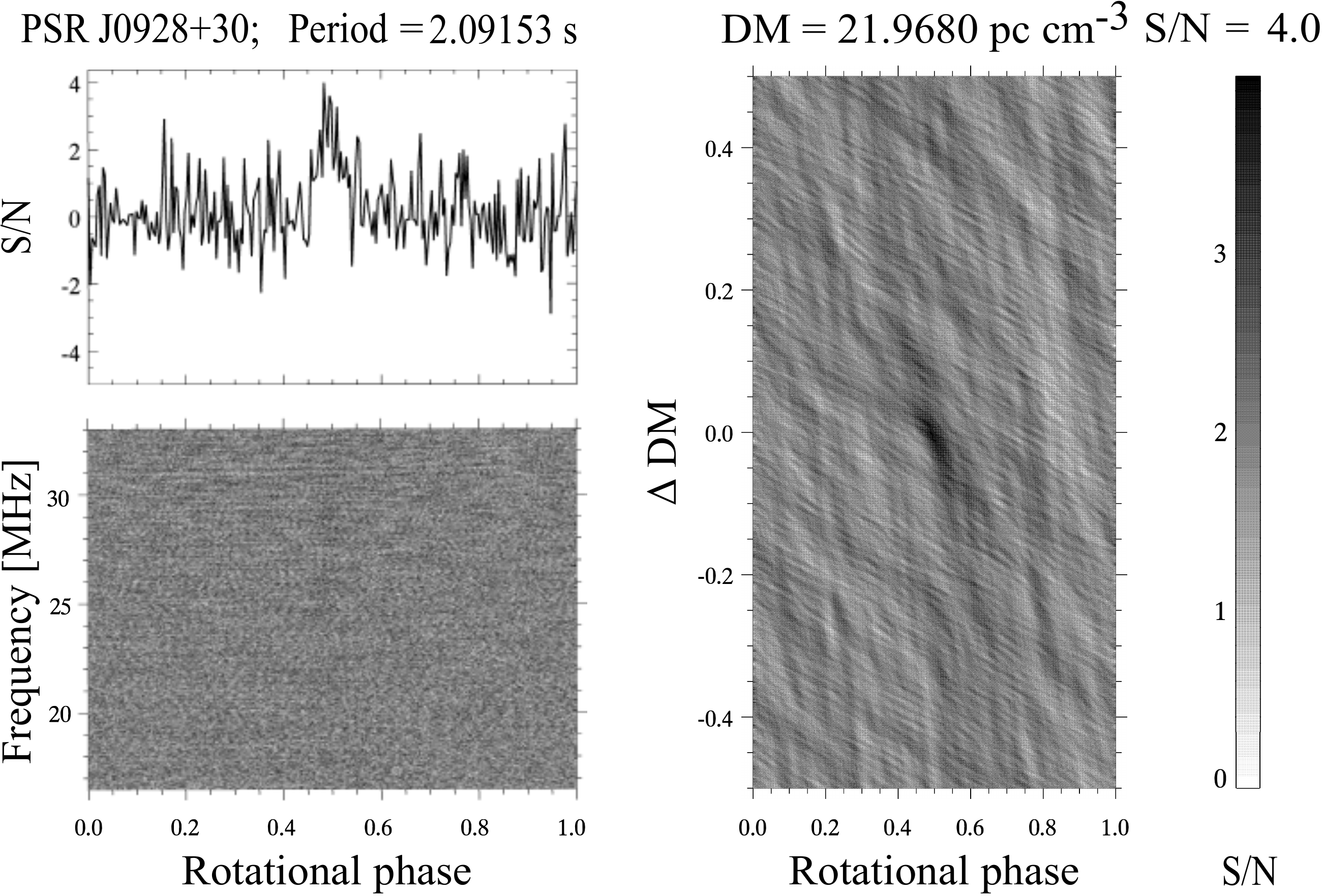}
    \caption{Detection of PSR\,J0928+30 on February 9, 2021 with the refined rotation period $P$\,=\,2.09153\,s.}
    \label{fig:PSRJ0928+30.pdf}
\end{figure}

{\itshape {PSR\,J0928+30}}. \citet{Sanidas2019} give DM\,=\,21.95$\pm$$0.09$\,pc\,cm$^{-3}$ and $P$\,=\,2.0915\,s for this pulsar. With the optimal time of observations and the minimal RFI level, we found the average pulse profile at the detection limit with S/N\,=\,4 (Fig.~\ref{fig:PSRJ0928+30.pdf}). However, it allowed us to measure the period $P$\,=\,2.09153\,s, and the DM\,=\,21.968$\pm$$0.020$\,pc\,cm$^{-3}$. 

\begin{figure}
	\includegraphics[width=\columnwidth]{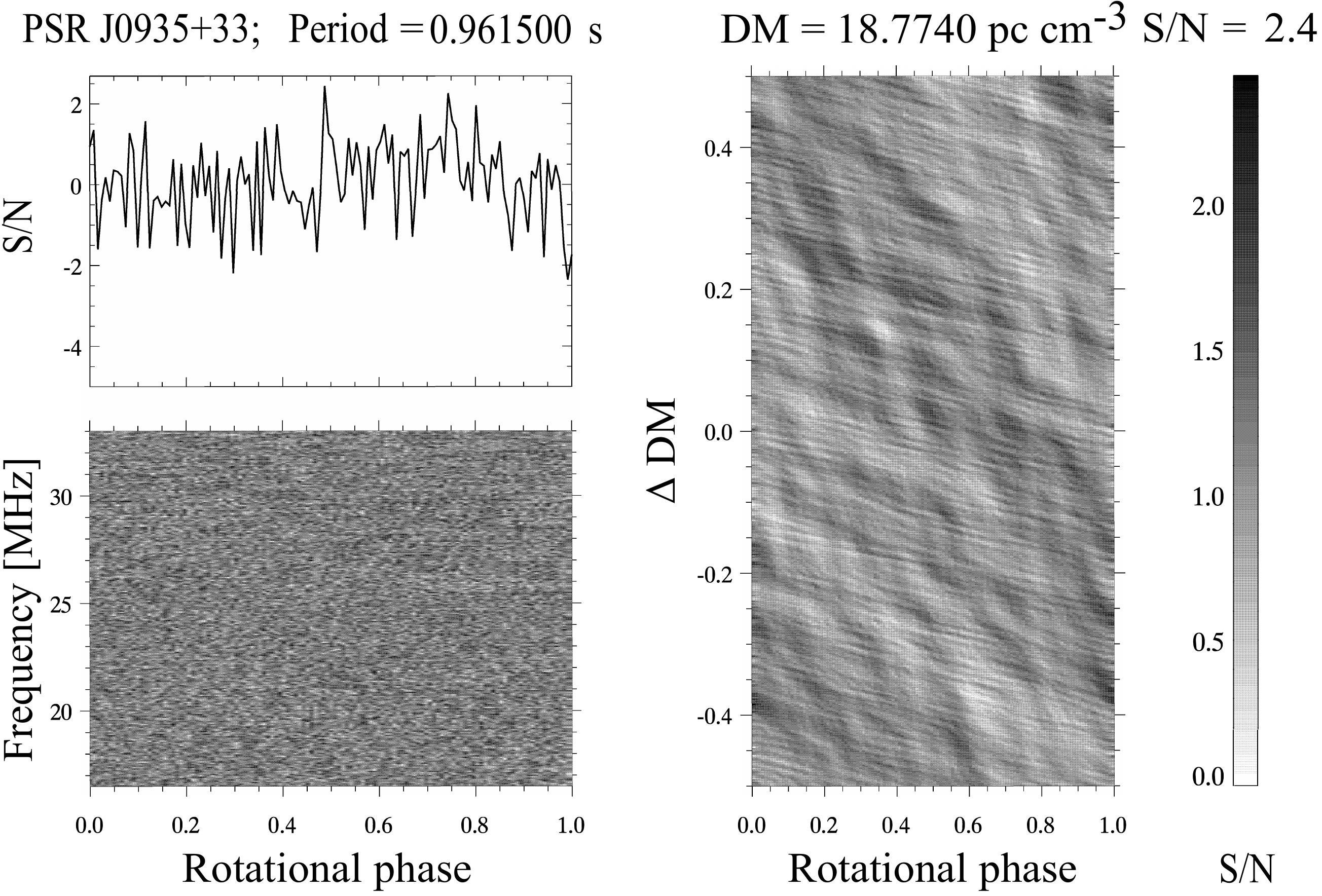}
    \caption{Unsuccessful detection of PSR\,J0935+33 on February 11, 2021 with rough DM\,=\,18.35$\pm$$0.06$\,pc\,cm$^{-3}$ and period $P$\,=\,$0.9615$\,s values.}
    \label{fig:PSRJ0935+33_1.pdf}
\end{figure}

\begin{figure}
	\includegraphics[width=\columnwidth]{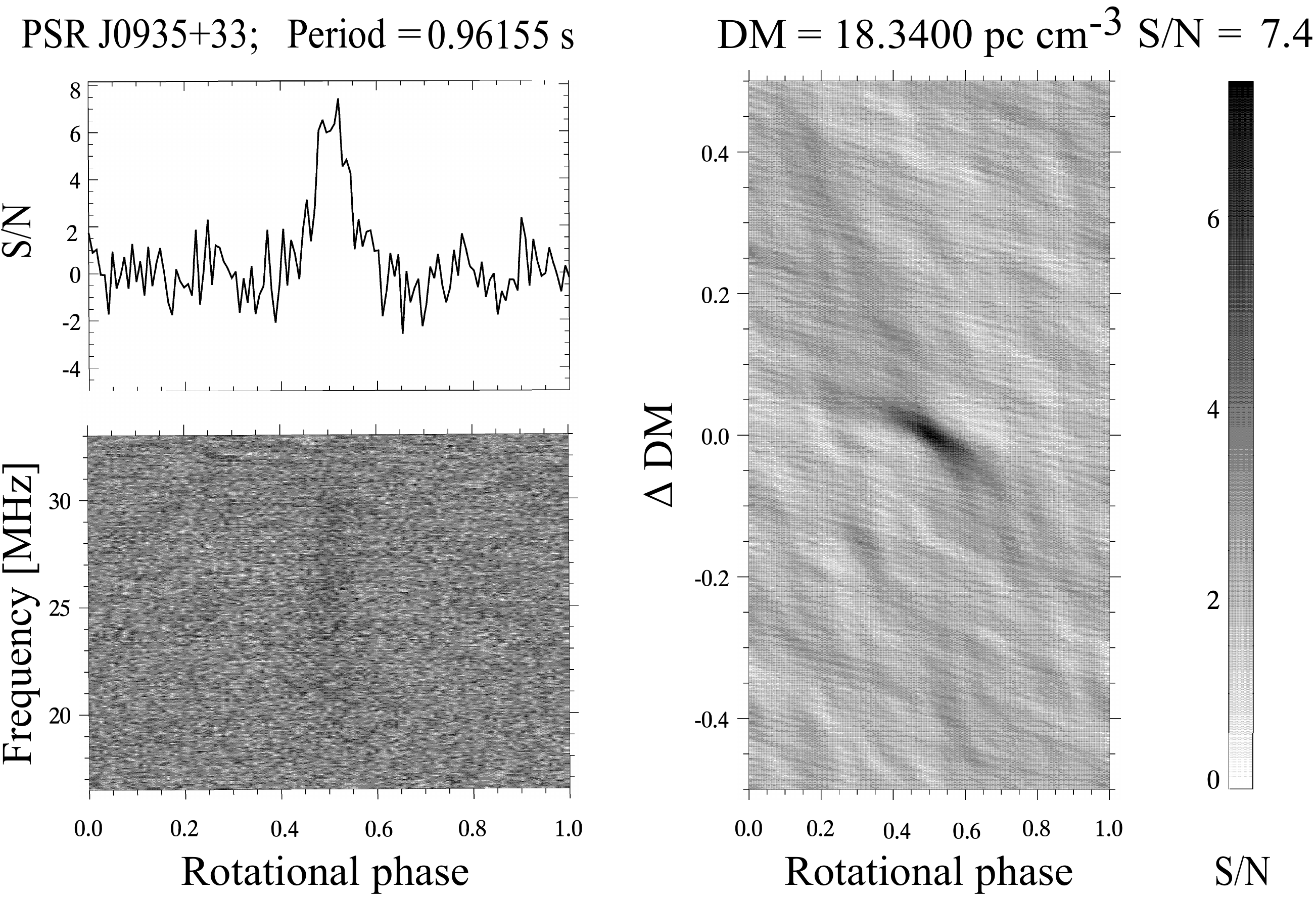}
    \caption{Detection of PSR\,J0935+33 on February 11, 2021 with the refined rotation period $P$\,=\,$0.96155$\,s.}
    \label{fig:PSRJ0935+33_2.pdf}
\end{figure}

{\itshape {PSR\,J0935+33}}. The latest measured parameters of this pulsar as for the early 2020 were $P$\,=\,0.9615\,s, and DM\,=\,18.35$\pm$$0.06$\,pc\,cm$^{-3}$ in \citet{Sanidas2019}. Using these parameters did not result in detection (see Fig.~\ref{fig:PSRJ0935+33_1.pdf}), but the period value tuning to the value $P$\,=\,0.96155\,s allowed us to find reliably the average pulse profile (Fig.~\ref{fig:PSRJ0935+33_2.pdf}) and to measure the DM\,=\,18.34$\pm$$0.02$\,pc\,cm$^{-3}$.

\begin{figure}
	\includegraphics[width=\columnwidth]{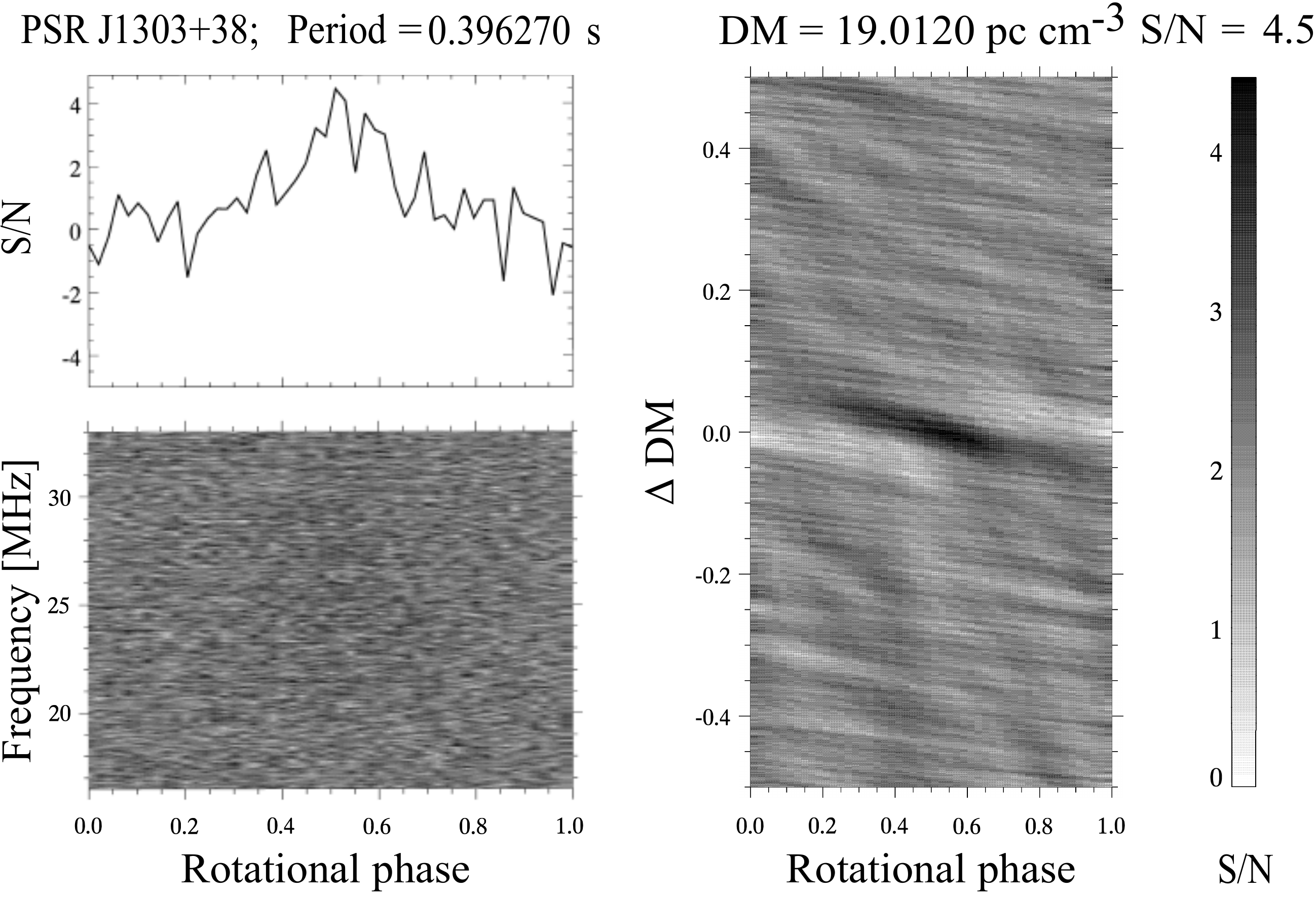}
    \caption{Detection of PSR\,J1303+38 on December 26, 2020 with the refined rotation period $P$\,=\,0.39627\,s.}
    \label{fig:PSRJ1303+38.pdf}
\end{figure}

{\itshape {PSR\,J1303+38}}. \citet{Sanidas2019} give the $P$\,=\,0.3963\,s, and DM\,=\,19.000$\pm$$0.009$\,pc\,cm$^{-3}$ for this pulsar, but using these parameters also did not result in a detection, so we had to tune the period value. The rotation period $P$\,=\,0.39627\,s gave a weak average pulse profile near the detection limit (Fig.~\ref{fig:PSRJ1303+38.pdf}) which allowed us to measure the DM\,=\,19.012$\pm$$0.020$\,pc\,cm$^{-3}$.

{\itshape {PSR\,J1426+52}}. We have already shown this pulsar in Section \ref{sec:Instrumentation and observations} (see Fig.~\ref{fig:PSRJ1426+52_2.pdf}) as an example of period refinement, we found its $P$\,=\,0.995866\,s and DM\,=\,25.332$\pm$$0.006$\,pc\,cm$^{-3}$.

\begin{figure}
	\includegraphics[width=\columnwidth]{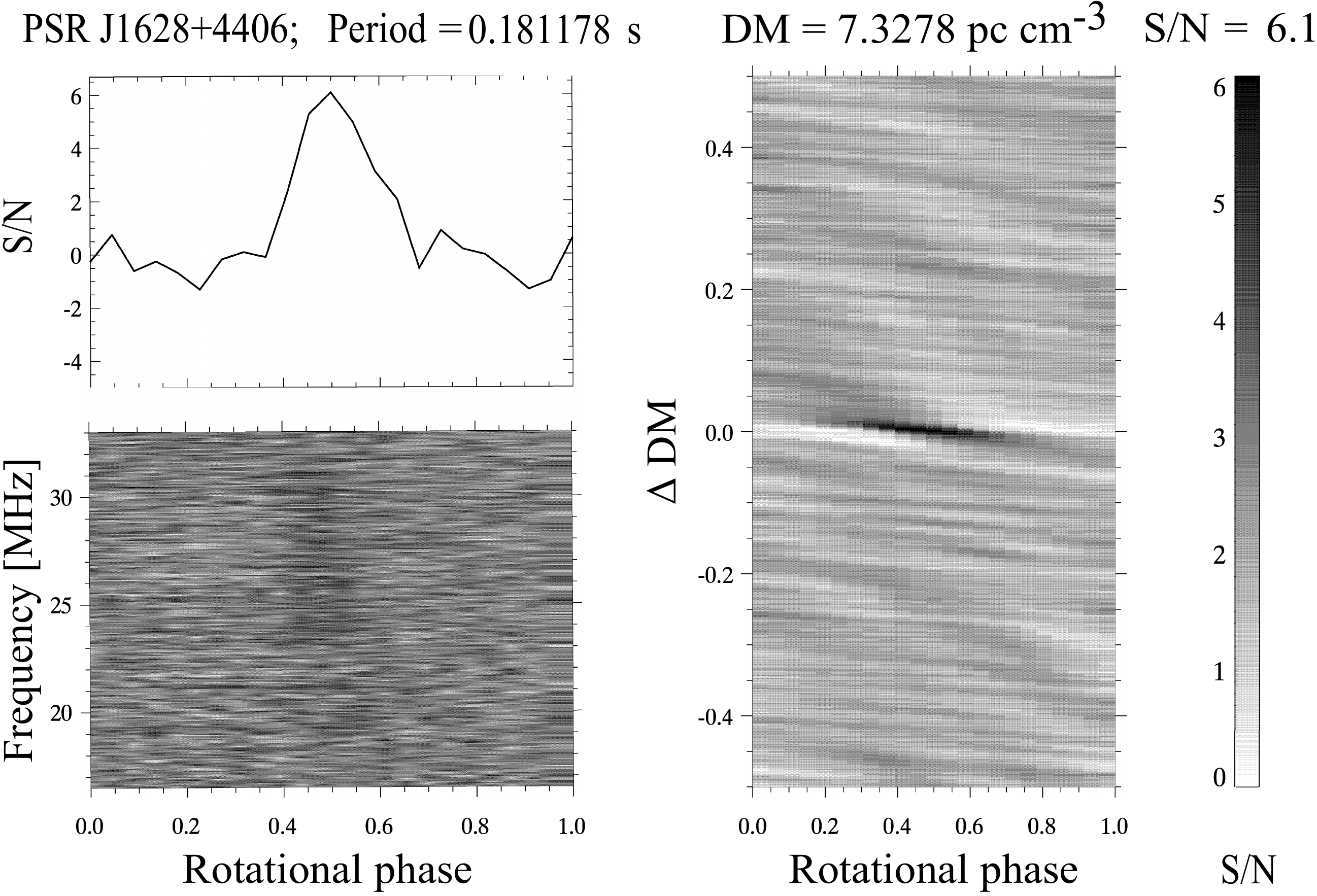}
    \caption{Detection of PSR\,J1628+4406 on May 26-27, 2020.}
    \label{fig:PSRJ1628+4406.pdf}
\end{figure}

{\itshape {PSR\,J1628+4406}}. The discovery of this pulsar was reported in \citet{Lynch2018} with DM\,=\,7.32981$\pm$$0.00002$\,pc\,cm$^{-3}$ and $P$\,=\,$0.18117848994714$\,s. This accurate period value allowed us to reliably record the average pulse profile with the S/N\,>\,6 (Fig.~\ref{fig:PSRJ1628+4406.pdf}) and to estimate its DM\,=\,7.3278$\pm$$0.0200$\,pc\,cm$^{-3}$. Since the period of PSR\,J1628+4406 is only 181\,ms, it is the fastest pulsar detected on decametre waves so far. 

\begin{figure}
	\includegraphics[width=\columnwidth]{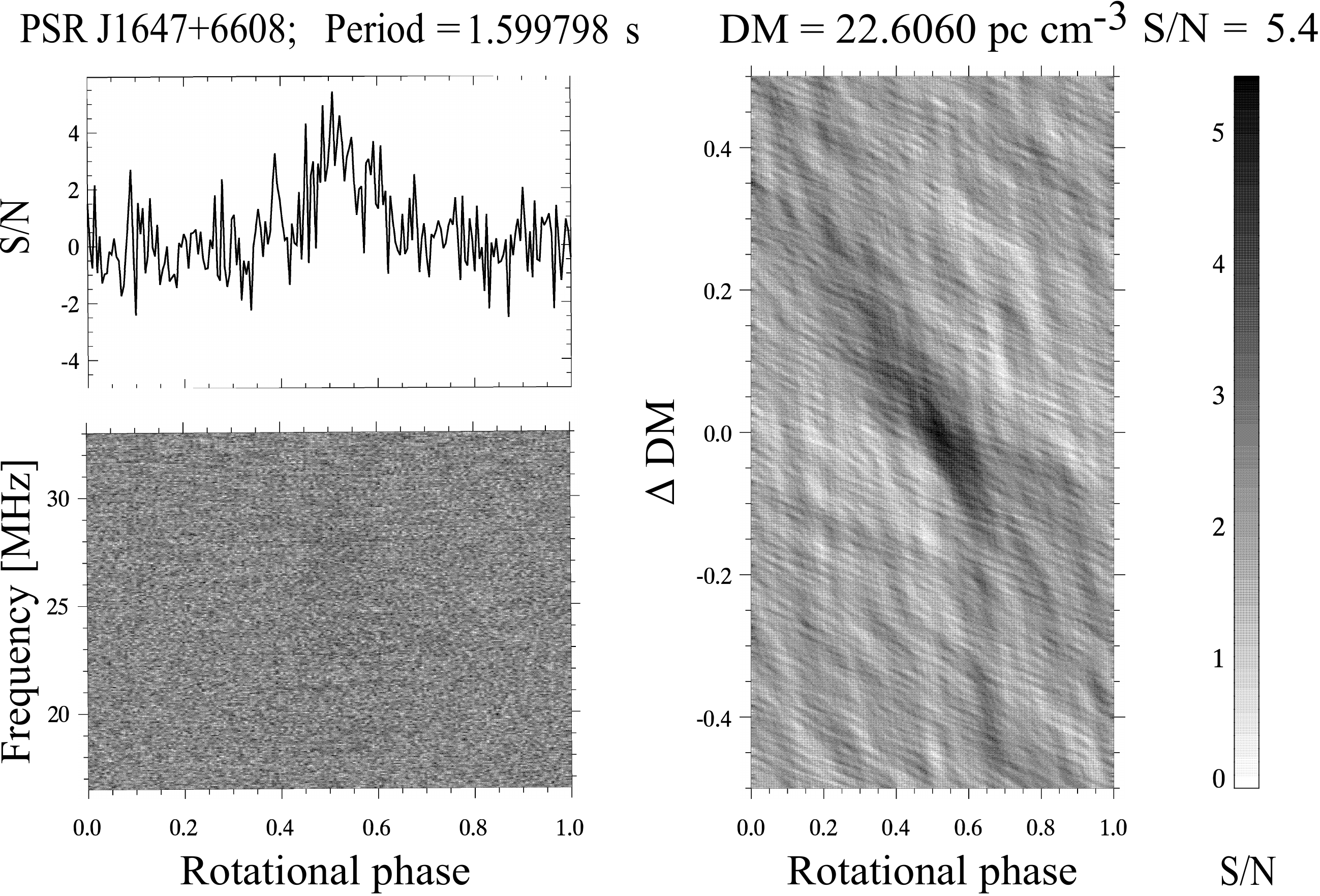}
    \caption{Detection of PSR\,J1647+6608 on September 8, 2020.}
    \label{fig:PSRJ1647+6608.pdf}
\end{figure}

{\itshape {PSR\,J1647+6608}}. The DM and rotation period of PSR\,J1647+6608 in \citet{Lynch2018} were estimated as 22.55$\pm$$0.07$\,pc\,cm$^{-3}$ and 1.59979837535\,s, respectively. We observed it in the evening of September 8, 2020 (culmination local time about 18h\,00m) and detected it quite reliably with S/N\,$>$\,9 (Fig.~\ref{fig:PSRJ1647+6608.pdf}) which allowed us to estimate its DM\,=\,22.606$\pm$$0.020$\,pc\,cm$^{-3}$.

\begin{figure}
	\includegraphics[width=\columnwidth]{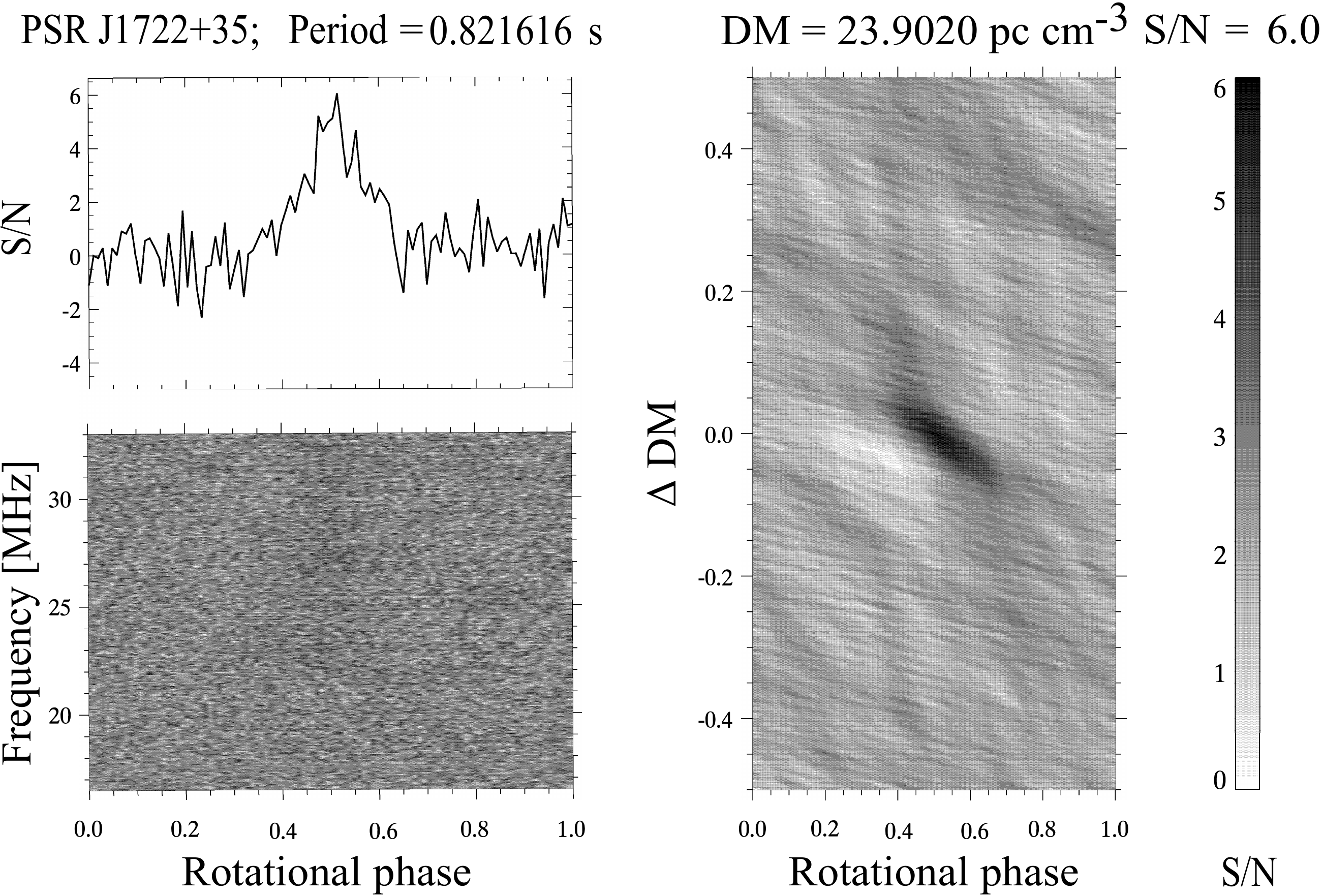}
    \caption{Detection of PSR\,J1722+35 on November 17, 2020 with the refined rotation period $P$\,=\,0.821616\,s.}
    \label{fig:PSRJ1722+35.pdf}
\end{figure}

{\itshape {PSR\,J1722+35}} was observed in the afternoon of November 17, 2020 (culmination local time about 13h\,00m). \citet{Sanidas2019} reports its DM\,=\,23.83$\pm$$0.06$\,pc\,cm$^{-3}$ and $P$\,=\,$0.8216$\,s. Despite the low accuracy of period measurements, we detected this pulsar. The average pulse profile and parameters obtained with $P$\,=\,$0.821616$\,s, DM\,=\,23.902$\pm$$0.020$\,pc\,cm$^{-3}$), are shown in Fig.~\ref{fig:PSRJ1722+35.pdf}.

\begin{figure}
	\includegraphics[width=\columnwidth]{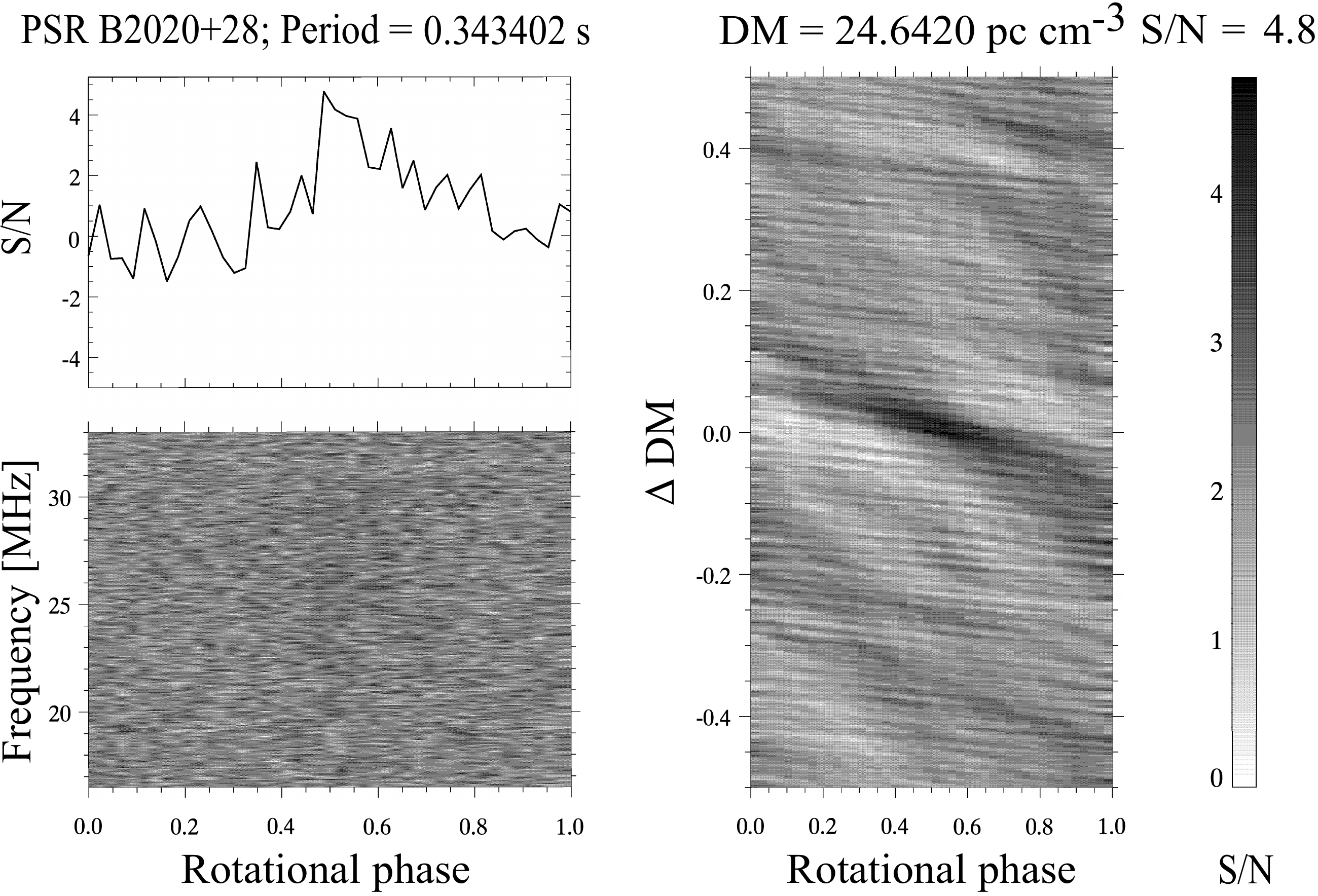}
    \caption{Detection of PSR\,B2020+28 on June 1, 2020.}
    \label{fig:PSRB2020+28.pdf}
\end{figure}

{\itshape {PSR\,B2020+28}}. In \citet{Bilous2016} and \citet{Hobbs2004} its DM\,=\,24.63109$\pm$$0.00018$\,pc\,cm$^{-3}$ and rotation period $P$\,=\,0.343402158\,s are given respectively. This pulsar like PSR\,J1628+4406 has a fairly short period of 343\,ms, but it is still detectable at low frequencies (Fig.~\ref{fig:PSRB2020+28.pdf}). Our estimate of its DM\,=\,24.642$\pm$$0.020$\,pc\,cm$^{-3}$.

\begin{figure}
	\includegraphics[width=\columnwidth]{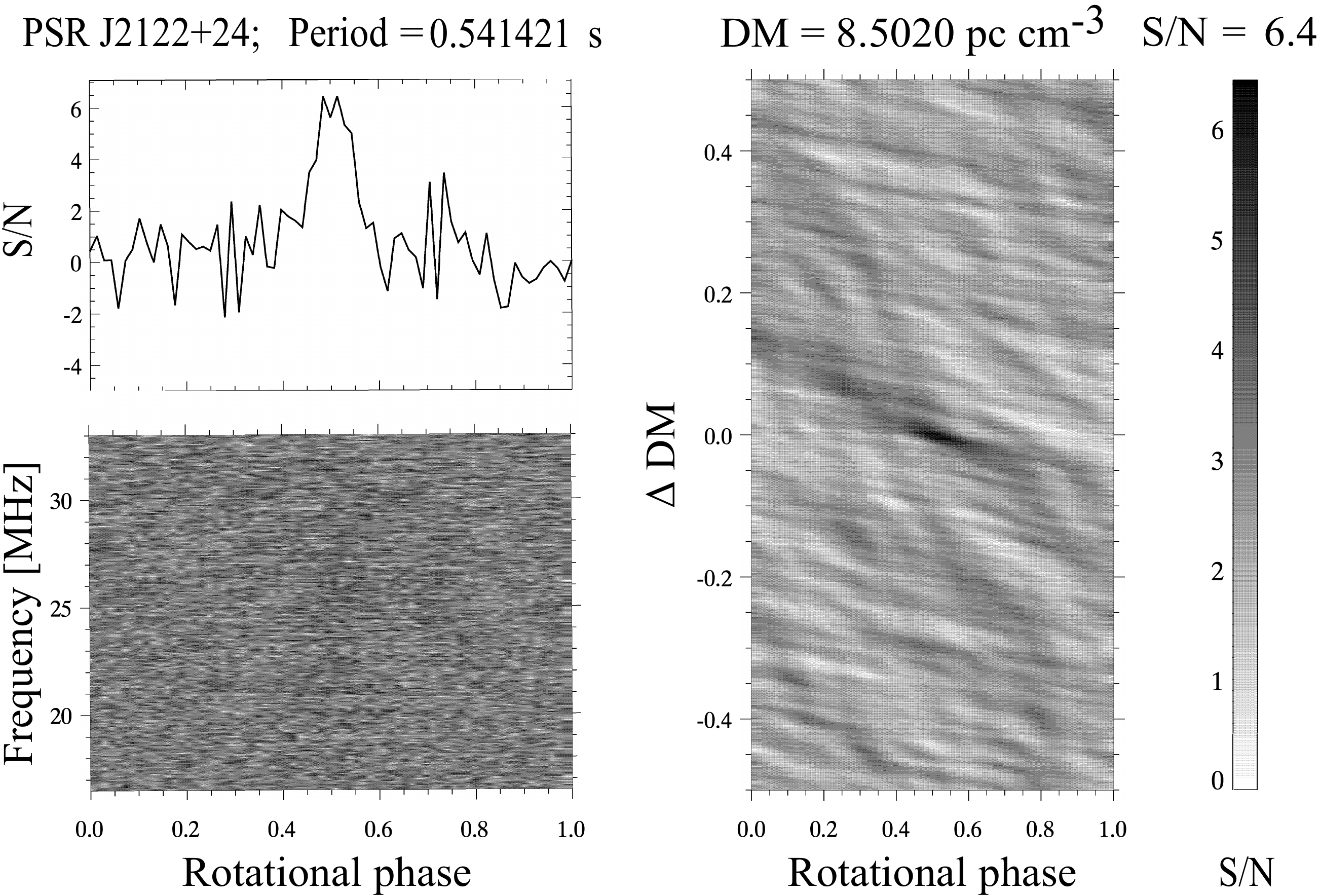}
    \caption{Detection of PSR\,J2122+24 on November 21, 2020.}
    \label{fig:PSRJ2122+24.pdf}
\end{figure}

{\itshape {PSR\,J2122+24}} was observed in September and November 2020. Its parameters in \citet{Tan2020} are: DM\,=\,8.500$\pm$$0.005$\,pc\,cm$^{-3}$ and $P$\,=\,0.54142115903\,s. It was best detected on November 21, 2020 when its culmination was at approximately 16:50 local time, the average pulse profile is shown in Fig.~\ref{fig:PSRJ2122+24.pdf}. Our measured DM\,=\,8.502$\pm$$0.020$\,pc\,cm$^{-3}$. 

\begin{figure}
	\includegraphics[width=\columnwidth]{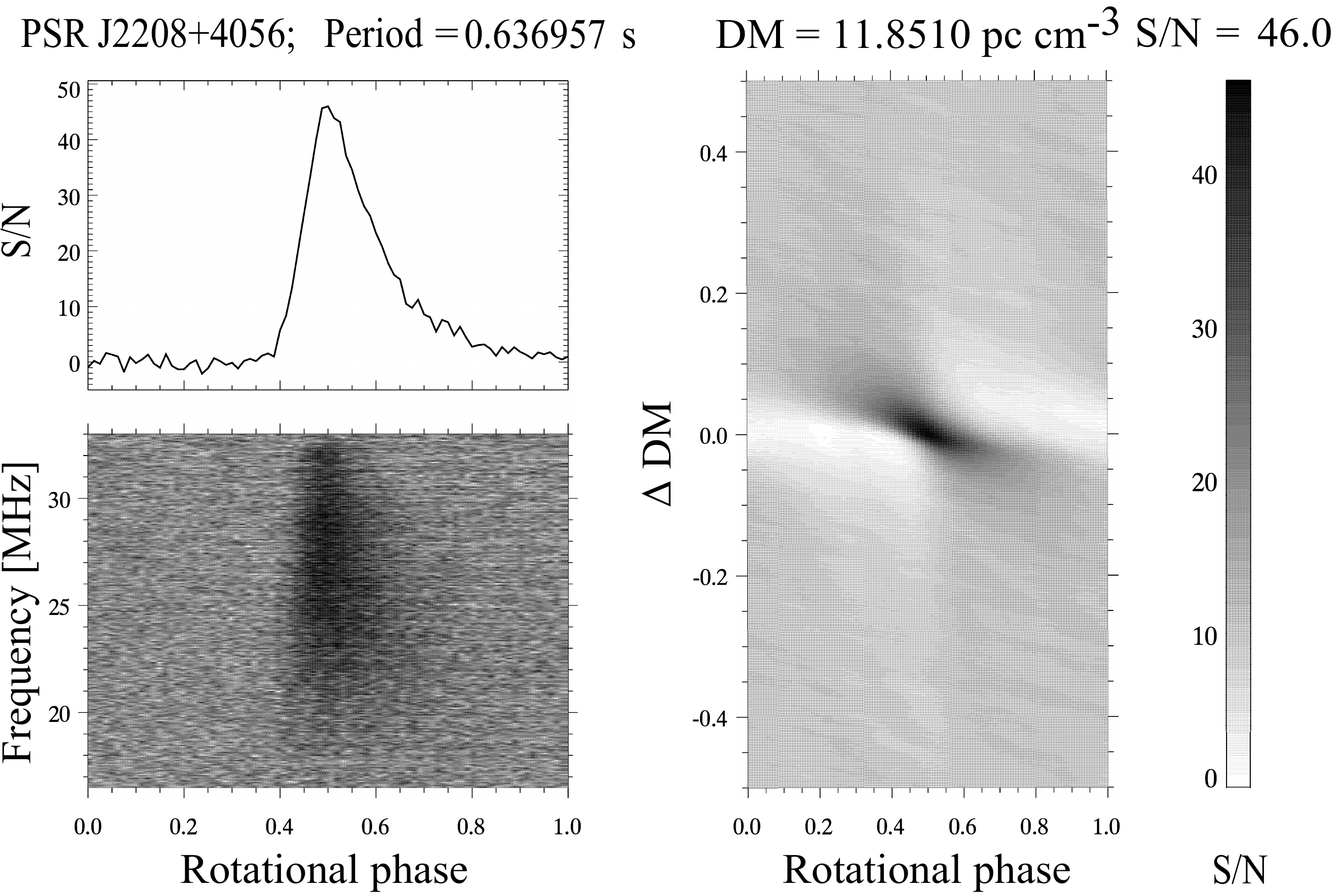}
    \caption{Detection of PSR\,J2208+4056 on February 14, 2020.}
    \label{fig:PSRJ2208+4056.pdf}
\end{figure}

{\itshape {PSR\,J2208+4056}} is the source with the highest S/N\,=\,46 obtained during the census. In \citet{Lynch2018} the estimates of its parameters are: DM\,=\,11.837$\pm$$0.009$\,pc\,cm$^{-3}$, $P$\,=\,0.63695739361\,s. We observed it on February 14, 2020 in the afternoon (culmination local time about noon). Despite the daytime observations, it was detected reliably (Fig.~\ref{fig:PSRJ2208+4056.pdf}) and our estimate of its DM\,=\,11.851$\pm$$0.006$\,pc\,cm$^{-3}$, which is close to the previous estimates. The scattering time constant of this pulsar is $\tau_{sc}$\,=\,68$\pm$8\,ms.

\begin{figure}
	\includegraphics[width=\columnwidth]{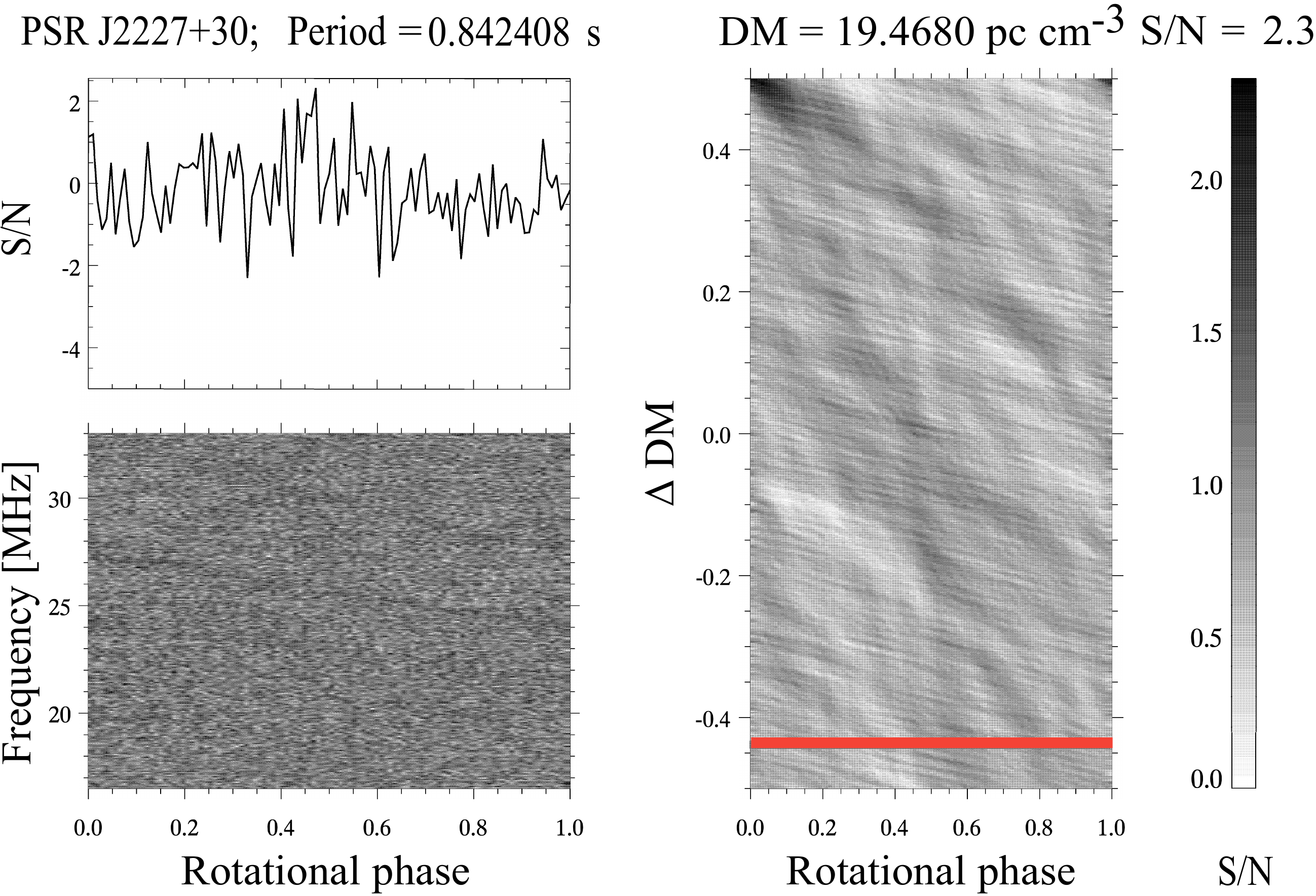}
    \caption{Unsuccessful detection of PSR\,J2227+30 on February 9, 2020 with the DM value from \citet{Tyulbashev2018} (DM from the paper is indicated with the red line).}
    \label{fig:PSRJ2227+30_1_plus_line.pdf}
\end{figure}

\begin{figure}
	\includegraphics[width=\columnwidth]{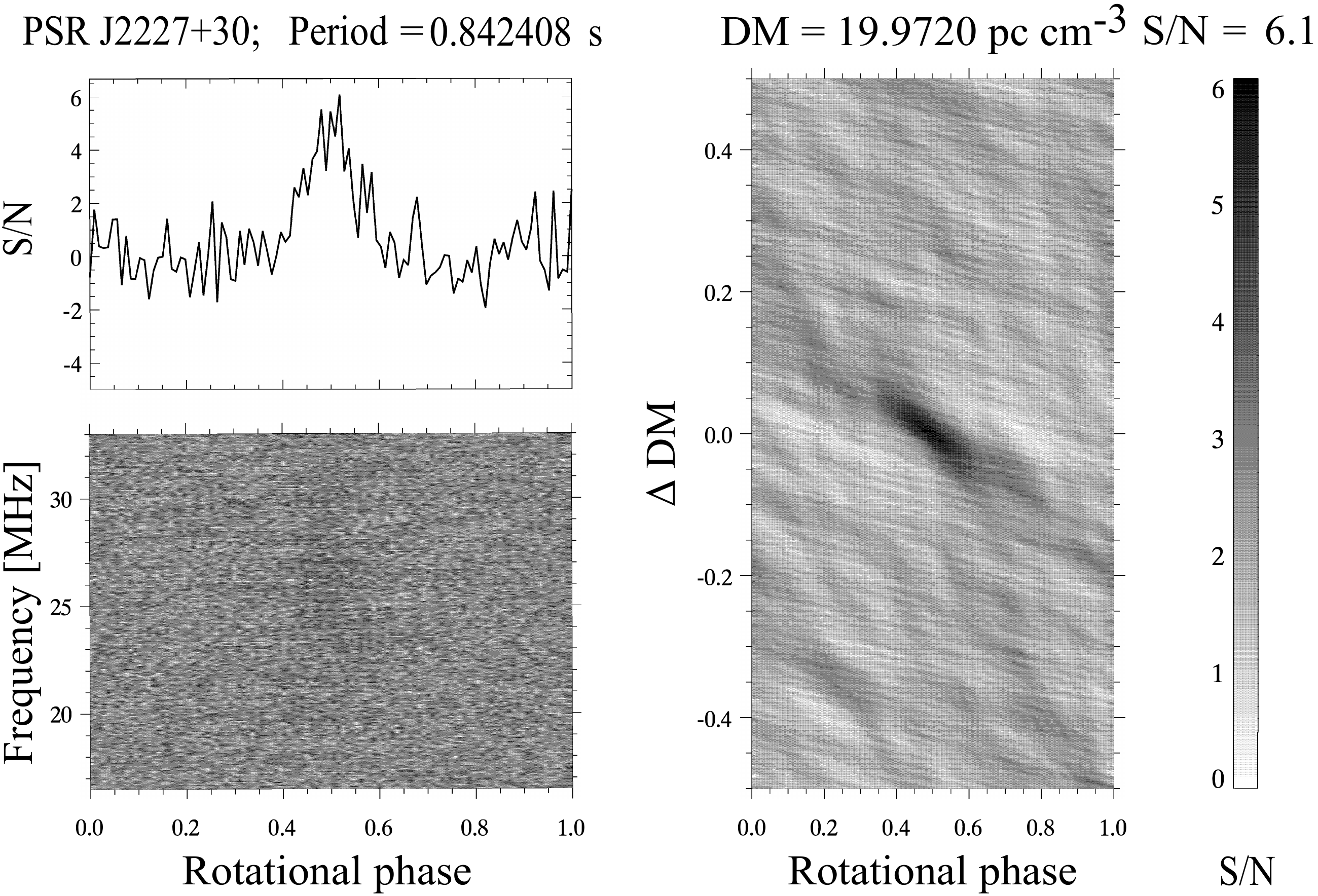}
    \caption{Detection of PSR\,J2227+30 on February 9, 2020 with the refined DM value.}
    \label{fig:PSRJ2227+30_2.pdf}
\end{figure}

{\itshape {PSR\,J2227+30}}. We observed it on February 9, 2020 in the afternoon (culmination was about 12:40 local time). The DM\,=\,19$\pm$$2$\,pc\,cm$^{-3}$ is given in \citet{Tyulbashev2018} and the period $P$\,=\,0.842408\,s we took from \citet{Camilo1996}. Processing of data with these parameters did not give positive result which is well illustrated with the plots of DM value refining procedure in a DM range of $1$\,pc\,cm$^{-3}$ shown in Fig.~\ref{fig:PSRJ2227+30_1_plus_line.pdf}. However, in the right panel the upper left corner a part of a black oval is visible. Wide-range DM searching in our pipeline allowed to detect PSR\,J2227+30 quite reliably (Fig.~\ref{fig:PSRJ2227+30_2.pdf}). Our measurements gave the DM\,=\,19.972$\pm$$0.020$\,pc\,cm$^{-3}$, which is $0.972$\,pc\,cm$^{-3}$ bigger than specified in \citet{Tyulbashev2018}, but within its error interval. In a situation with a high influence of RFI and a narrow relative observation bandwidth, the accuracy of the DM value measurement in low-frequency studies (at least in the decametre and metre wave ranges) is much lower.

\begin{figure}
	\includegraphics[width=\columnwidth]{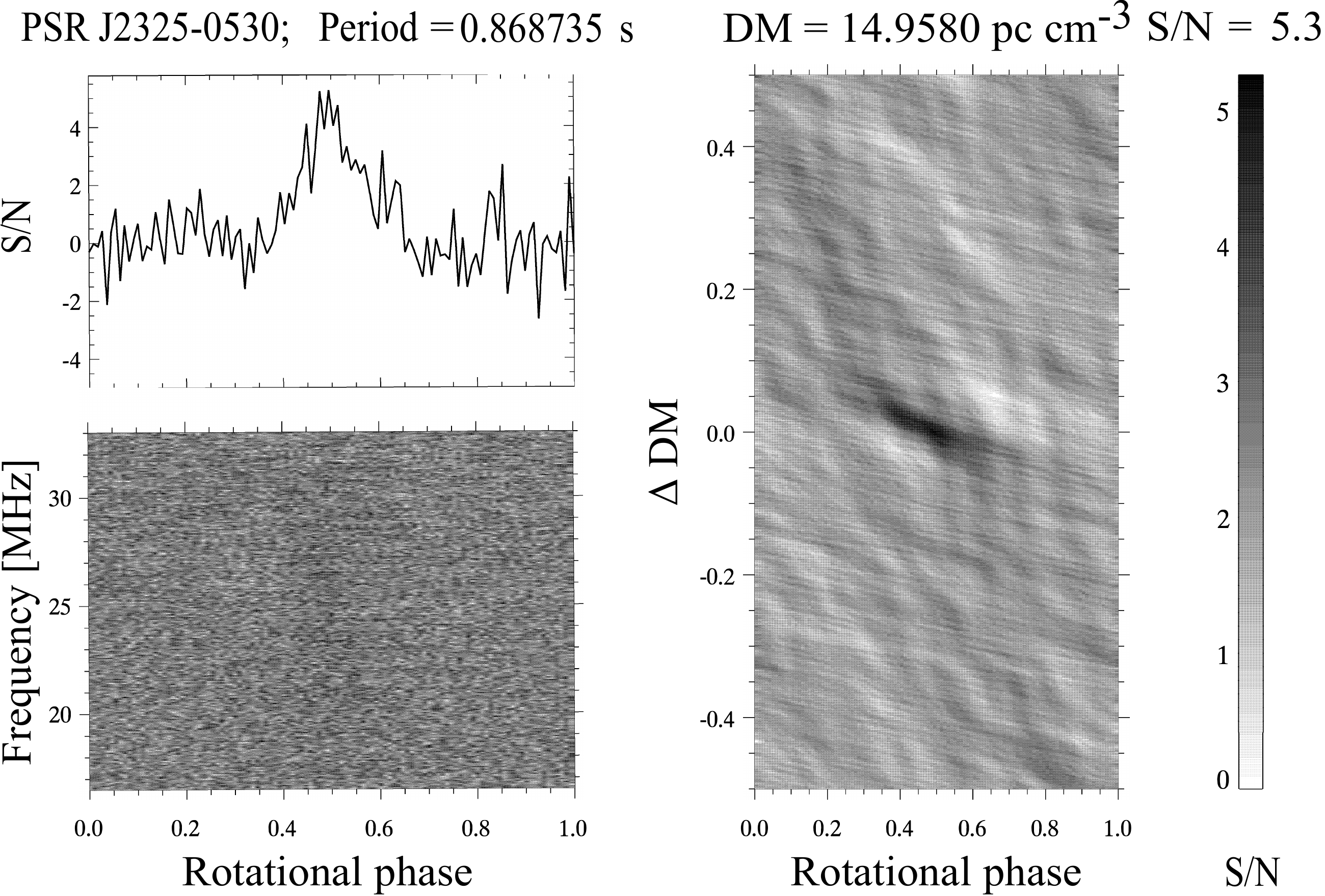}
    \caption{Detection of PSR\,J2325-0530 on September 7, 2020.}
    \label{fig:PSRJ2325-0530.pdf}
\end{figure}

{\itshape {PSR\,J2325-0530}} was observed in February and September 2020. In \citet{Karako-Argaman2015} its parameters are estimated as: DM\,=\,14.966$\pm$$0.007$\,pc\,cm$^{-3}$, $P$\,=\,0.868735115026\,s. In the second observation session on September 7, 2020, it was detected most reliably (Fig.~\ref{fig:PSRJ2325-0530.pdf}) because of night-time observations (culmination about 0:50 local time). The obtained DM\,=\,14.958$\pm$$0.020$\,pc\,cm$^{-3}$ practically equals to the value given in \citet{Karako-Argaman2015}.

{\itshape {PSR\,J2336-01}} was observed on February 15, 2020 in the afternoon (culmination about 13:30 local time). Its parameters in \citet{Sanidas2019} were not accurate enough: DM\,=\,19.60$\pm$$0.09$\,pc\,cm$^{-3}$, rotation period $P$\,=\,1.0298\,s. Although we detected the radiation of this pulsar even using this period value, we refined this parameter to $P$\,=\,1.0298454\,s, which resulted in measurement of DM\,=\,19.624$\pm$$0.020$\,pc\,cm$^{-3}$ (Fig.~\ref{fig:PSRJ2325-0530.pdf}).

\begin{figure}
	\includegraphics[width=\columnwidth]{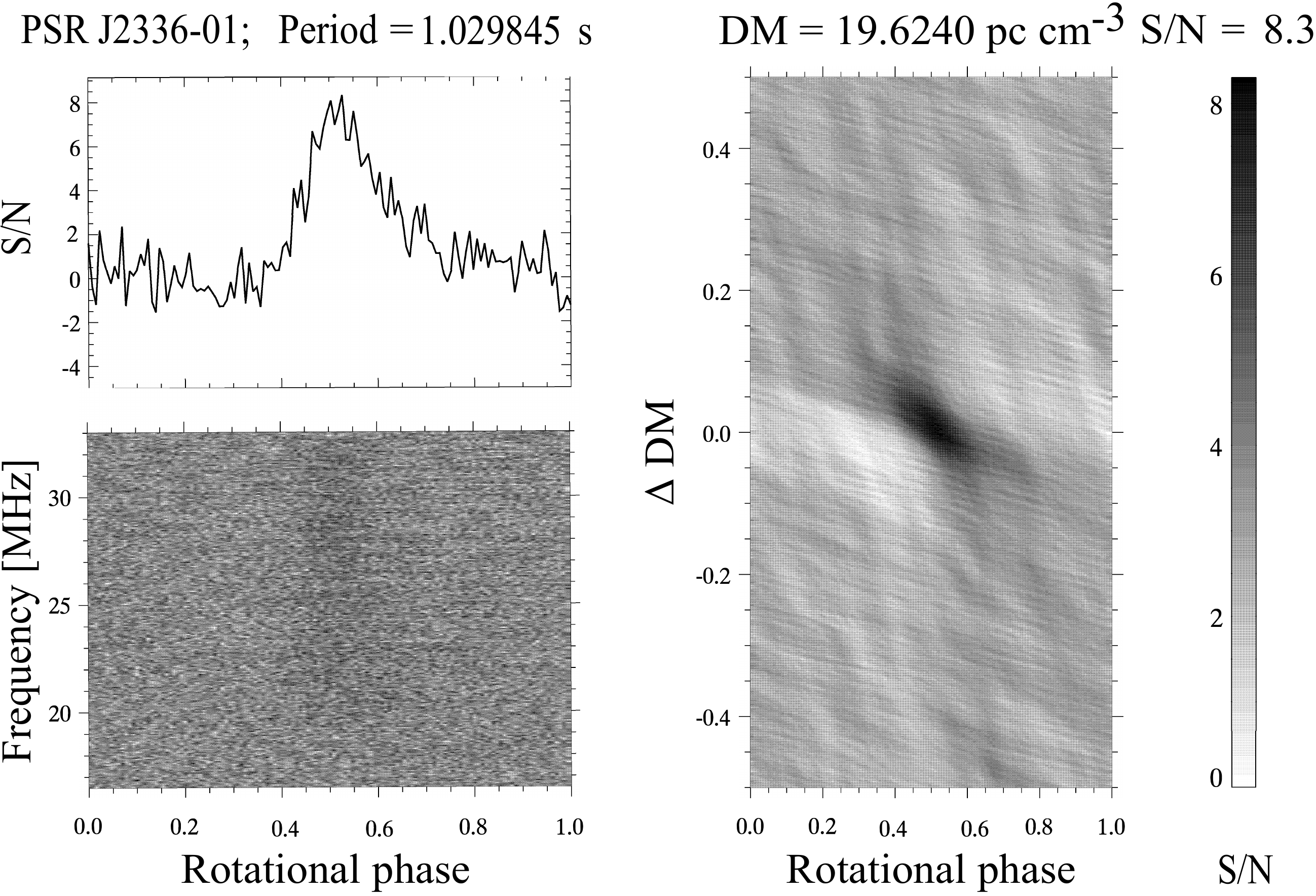}
    \caption{Detection of PSR\,J2336-01 on February 15, 2020 with the specified period $P$\,=\,1.0298454\,s.}
    \label{fig:PSRJ2336-01.pdf}
\end{figure}

\subsection{Analysis, discussion and prospects}
\label{sec:Analysis, discussion and prospects} 

In order to demonstrate more clearly our refinement of the DM of a number of pulsars, we have summarized them all in Table~\ref{tab:table1}. The sixth column shows the target DM values for each pulsar obtained at higher frequencies, the seventh one – our DM values.

Note the pulsars detected with high enough S/N in which the DMs measured in this work differ significantly from those measured at high frequencies. For pulsars PSR\,J0121+14, PSR\,J0811+37, PSR\,J1426+52, PSR\,J1722+35, PSR\,B2020+28, PSR\,J2208+4056, PSR\,J2227+30, PSR\,J2325-0530, PSR\,J2336-01 the DM accuracy obtained in this work is rather high, so the new DM values for these pulsars are quite reliable. Flux densities and FWHM for each pulsar are also indicated in the table. The estimation error of flux densities as in \citet{Zakharenko2013} equals to 50\%. 

\begin{table*}
    \centering
    \caption{Flux densities, rough (from previous works) and our refined  dispersion measures and other parameters for twenty pulsars detected in decameter wave range in this work.}
    \label{tab:table1}
    \begin{tabular}{ccccccccc} 
		    \hline
		    N & PSR~name & Gl\,[$^{\circ}$] & Gb\,[$^{\circ}$] & JD & «Aim»~DM\,[\,pc\,cm$^{-3}$] & Refined DM\,[\,pc\,cm$^{-3}$] & Flux\,[mJy] & FWHM\,[ms]\\
		    \hline
		    1 & J0121+1 & 134.023 & -47.946 & 2458890.0000000 & 17.77(9) & 17.690(20) & 22$\pm$11 & 78$\pm$8\\
		    2 & B0301+19 & 161.135 & -33.27 & 2458885.0868056 & 15.65677(35) & 15.659(20) & 4.7$\pm$2.4 & \\
		    3 & J0317+13 & 168.763 & -36.04 & 2458896.0666782 & 12.90(4) & 13.148(20) & 6.1$\pm$3.1 & \\
	    	4 & J0454+45 & 160.719 & 1.212 & 2459254.1541898 & 20.834(2) & 20.854(20) & 27$\pm$14 & \\
		    5 & J0613+3731 & 175.336 & 9.239 & 2459172.4285532 & 18.990(12) & 18.986(20) & 20$\pm$10 & 72$\pm$8\\
		    6 & J0811+37 & 183.676 & 31.217 & 2458889.2889005 & 16.95(11) & 17.124(20) & 5.8$\pm$2.9 & 70$\pm$8\\
	    	7 & B0917+63 & 151.431 & 40.725 & 2458998.0402894 & 13.15423(18) & 13.156(10) & 13$\pm$7 & 98$\pm$8\\
		    8 & J0928+30 & 195.838 & 45.915 & 2459255.3506944 & 21.95(9) & 21.968(20) & 3.6$\pm$1.8 & \\
		    9 & J0935+33 & 192.376 & 47.511 & 2459257.3396065 & 18.35(6) & 18.34(20) & 6.6$\pm$3.3 & 73$\pm$8\\
		    10 & J1303+38 & 111.027 & 78.636 & 2459209.6177662 & 19.000(9) & 19.012(20) & 9$\pm$4.5 & \\
	    	11 & J1426+52 & 93.88 & 59.233 & 2458885.5596644 & 25.37(2) & 25.332(6) & 34$\pm$17 & 88$\pm$8\\
		    12 & J1628+4406 & 69.239 & 43.616 & 2458996.3402778 & 7.32981(2) & 7.327(200) & 6.9$\pm$3.5 & \\
		    13 & J1647+6608 & 97.18 & 37.034 & 2459101.0666782 & 22.55(7) & 22.606(20) & 15$\pm$8 & \\
	    	14 & J1722+35 & 59.218 & 32.651 & 2459170.9000116 & 23.83(6) & 23.876(20) & 18$\pm$9 & \\
	    	15 & B2020+28 & 68.863 & -4.671 & 2459001.5048148 & 24.63109(18) & 24.642(20) & 24$\pm$12 & \\
	    	16 & J2122+24 & 73.79 & -17.973 & 2459175.0562616 & 8.500(5) & 8.502(20) & 24$\pm$12 & \\
	    	17 & J2208+4056 & 92.574 & -12.112 & 2458893.8576505 & 11.837(9) & 11.851(6) & 214$\pm$107 & 101$\pm$8\\
	    	18 & J2227+30 & 89.656 & -22.815 & 2458888.8847222 & 19(2) & 19.972(20) & 16$\pm$8 & \\
	    	19 & J2325-0530 & 75.579 & -60.2 & 2459100.3459259 & 14.966(7) & 14.958(20) & 19$\pm$10 & \\
	    	20 & J2336-01 & 84.424 & -59.008 & 2459171.1598495 & 19.60(9) & 19.624(20) & 47$\pm$24 & 202$\pm$16\\
    \hline
	\end{tabular}
\end{table*}

For six reliably detected pulsars, we estimated their scattering time constant (see Table~\ref{tab:table2}). The scattering time constant of all detected pulsars is much less than their period and it is the main feature which allows us to distinguish the on-pulse component against the background of off-pulse. If $\tau_{sc}/P$ ratio is close to 1, the detection is significantly more difficult, because in this case the differences between the on-pulse and off-pulse components levels depend on the scattering of the pulse. So far, we can state that all pulsars with DM values, at least, up to $30$\,pc\,cm$^{-3}$ should be well detectable at decameter waves because the on-pulse and off-pulse components are well distinguishable in case of enough sensitivity.

\begin{table*}
	\centering
	\caption{Parameters of pulsars with estimated the scattering time constant $\tau_{sc}$ in decameter wave range.}
	\label{tab:table2}
	\begin{tabular}{ccccccc} 
		\hline
		N & PSR\,name & DM\,[\,pc\,cm$^{-3}$] & S/N & $\tau_{sc}$\,[s] & $P$\,[s] & $\tau_{sc}/P$\\ 
		\hline
		1 & J0121+14 & 17.690(20) & 16.2 & 0.051(16) & 1.388993 & 0.0367\\
		2 & J0613+3731 & 18.986(20) & 26.2 & 0.030(10) & 0.619198181 & 0.0484\\
		3 & J0811+37 & 17.124(20) & 26.4 & 0.033(10) & 1.24827 & 0.0264\\
		4 & B0917+63 & 13.156(10) & 14.8 & 0.015(8) & 1.567994018 & 0.00957\\
		5 & J1426+52 & 25.332(6) & 18.1 & 0.053(16) & 0.995866 & 0.0532\\
		6 & J2208+4056 & 11.851(6) & 46 & 0.058(8) & 0.636957394 & 0.107\\
		\hline
	\end{tabular}
\end{table*}

In Table~\ref{tab:table3} the upper limits (at $4\sigma$ level) of flux densities of 102 not yet detected pulsars in assumption that they have 10\% duty cycle (pulse duration to pulsar period ratio) are given.

\tabcolsep=1mm
\renewcommand{\arraystretch}{0}

\begin{table}
\caption{The upper limits of flux density ($4\sigma$) of non-detected in this work pulsars.}
\label{tab:table3}
\begin{tabular}{|c|c|c|c|c|c|}
\hline N & Pulsar & Flux\,[mJy] & N & Pulsar & Flux\,[mJy]
\\ && (25\,MHz) &&& (25\,MHz)
 \\ \hline 1&J0006+1834&$\leq$ 39&52&J1332-03&$\leq$ 45
\\ \hline 2&J0011+08&$\leq$ 44&53&J1334+10&$\leq$ 50 \\
\hline 3&J0050+03&$\leq$ 45&54&J1336+33&$\leq$ 29 \\ \hline
4&J0054+66&$\leq$ 66&55&J1340+65&$\leq$ 48 \\ \hline
5&J0107+13&$\leq$ 41&56&J1404+1159&$\leq$ 53 \\ \hline
6&J0137+1654&$\leq$ 41&57&J1439+76&$\leq$ 36 \\ \hline
7&J0139+33&$\leq$ 40&58&J1501-0046&$\leq$ 60 \\ \hline
8&J0146+31&$\leq$ 39&59&J1502+28&$\leq$ 40 \\ \hline
9&J0152+0948&$\leq$ 44&60&BJ1503+2111&$\leq$ 56 \\ \hline
10&J0158+21&$\leq$ 40&61&JJ1518-0627&$\leq$ 68 \\ \hline
11&J0201+7005&$\leq$ 61&62&J1529+40&$\leq$ 31 \\ \hline
12&J0215+51&$\leq$ 88&63&J1536+17&$\leq$ 66\\ \hline
13&J0229+20&$\leq$ 40&64&J1538+2345&$\leq$ 52 \\ \hline
14&J0241+16&$\leq$ 43&65&J1549+2113&$\leq$ 60 \\ \hline
15&J0302+22&$\leq$ 40&66&J1555-0515&$\leq$ 73 \\ \hline
16&J0305+11&$\leq$ 44&67&J1603+18&$\leq$ 67 \\ \hline
17&J0332+79&$\leq$ 48&68&J1611-01&$\leq$ 75 \\ \hline
18&B0410+69&$\leq$ 41&69&J1612+2008&$\leq$ 64 \\ \hline
19&J0447-04&$\leq$ 48&70&J1623+58&$\leq$ 33 \\ \hline
20&J0459-0210&$\leq$ 47&71&J1657+33&$\leq$ 42 \\ \hline
21&J0517+2212&$\leq$ 51&72&J1703+00&$\leq$ 96 \\ \hline
22&B0609+37&$\leq$ 42&73&J1707+35&$\leq$ 41 \\ \hline
23&J0633+1746&$\leq$ 49&74&J1708+02&$\leq$ 101 \\ \hline
24&B0656+14&$\leq$ 46&75&J1715+46&$\leq$ 34 \\ \hline
25&B0655+64&$\leq$ 36&76&J1717+03&$\leq$ 107 \\ \hline
26&J0738+6904&$\leq$ 35&77&J1726+34&$\leq$ 66 \\ \hline
27&J0747+6646&$\leq$ 34&78&J1738+04&$\leq$ 62 \\ \hline
28&J0750+57&$\leq$ 32&79&J1740+1000&$\leq$ 106 \\ \hline
29&J0802-09&$\leq$ 54&80&J1800+5034&$\leq$ 38 \\ \hline
30&J0815+4611&$\leq$ 46&81&J1817-0743&$\leq$ 176 \\ \hline
31&B0820+02&$\leq$ 37&82&J1832+0029&$\leq$ 149 \\ \hline
32&J0857+33&$\leq$ 23&83&J1848+0647&$\leq$ 136 \\ \hline
33&J0943+2253&$\leq$ 25&84&J1850+15&$\leq$ 101 \\ \hline
34&J0944+4106&$\leq$ 21&85&J1910+56&$\leq$ 50 \\ \hline
35&J0947+2740&$\leq$ 24&86&J1917+0834&$\leq$ 127 \\ \hline
36&J0957-06&$\leq$ 39&87&B1916+14&$\leq$ 102 \\ \hline
37&J1000+08&$\leq$ 30&88&J1918+1541&$\leq$ 98 \\ \hline
38&J1011+18&$\leq$ 38&89&B1918+26&$\leq$ 73 \\ \hline
39&J1017+30&$\leq$ 24&90&J1929+16&$\leq$ 95 \\ \hline
40&J1046+0304&$\leq$ 33&91&J2015+2524&$\leq$ 75 \\ \hline
41&J1049+5822&$\leq$ 40&92&B2021+51&$\leq$ 63 \\ \hline
42&J1059+6459&$\leq$ 29&93&J2027+7502&$\leq$ 53 \\ \hline
43&J1110+58&$\leq$ 25&94&J2053+1718&$\leq$ 72 \\ \hline
44&J1132+25&$\leq$ 29&95&J2151+2315&$\leq$ 55 \\ \hline
45&J1226+00&$\leq$ 47&96&J2215+1538&$\leq$ 53 \\ \hline
46&J1235-02&$\leq$ 49&97&J2248-0101&$\leq$ 54 \\ \hline
47&J1242+39&$\leq$ 26&98&J2253+1516&$\leq$ 47 \\ \hline
48&J1243+17&$\leq$ 38&99&J2340+08&$\leq$ 46 \\ \hline
49&J1246+2253&$\leq$ 34&100&J2346-0609&$\leq$ 54 \\ \hline
50&J1313+0931&$\leq$ 48&101&J2354+40&$\leq$ 44 \\ \hline
51&J1320+67&$\leq$ 32&102&J2355+2246&$\leq$ 39 \\ \hline

\end{tabular}
\end{table}

\begin{figure*}
	\begin{minipage}[h]{0.48\linewidth}
    \center{\includegraphics[width=1\linewidth]{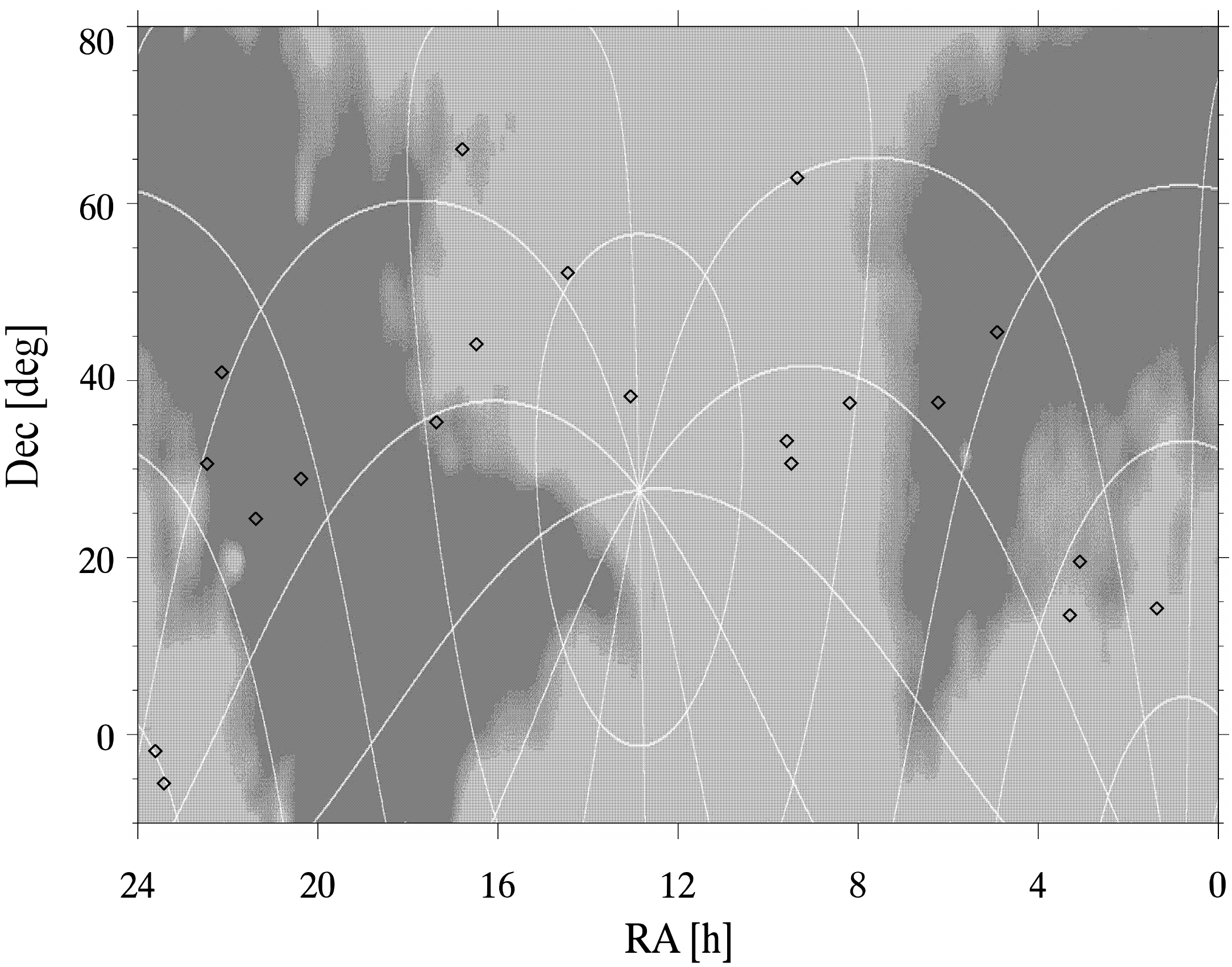} \\ a)}
    \end{minipage}
    \hfill
    \begin{minipage}[h]{0.48\linewidth}
    \center{\includegraphics[width=1\linewidth]{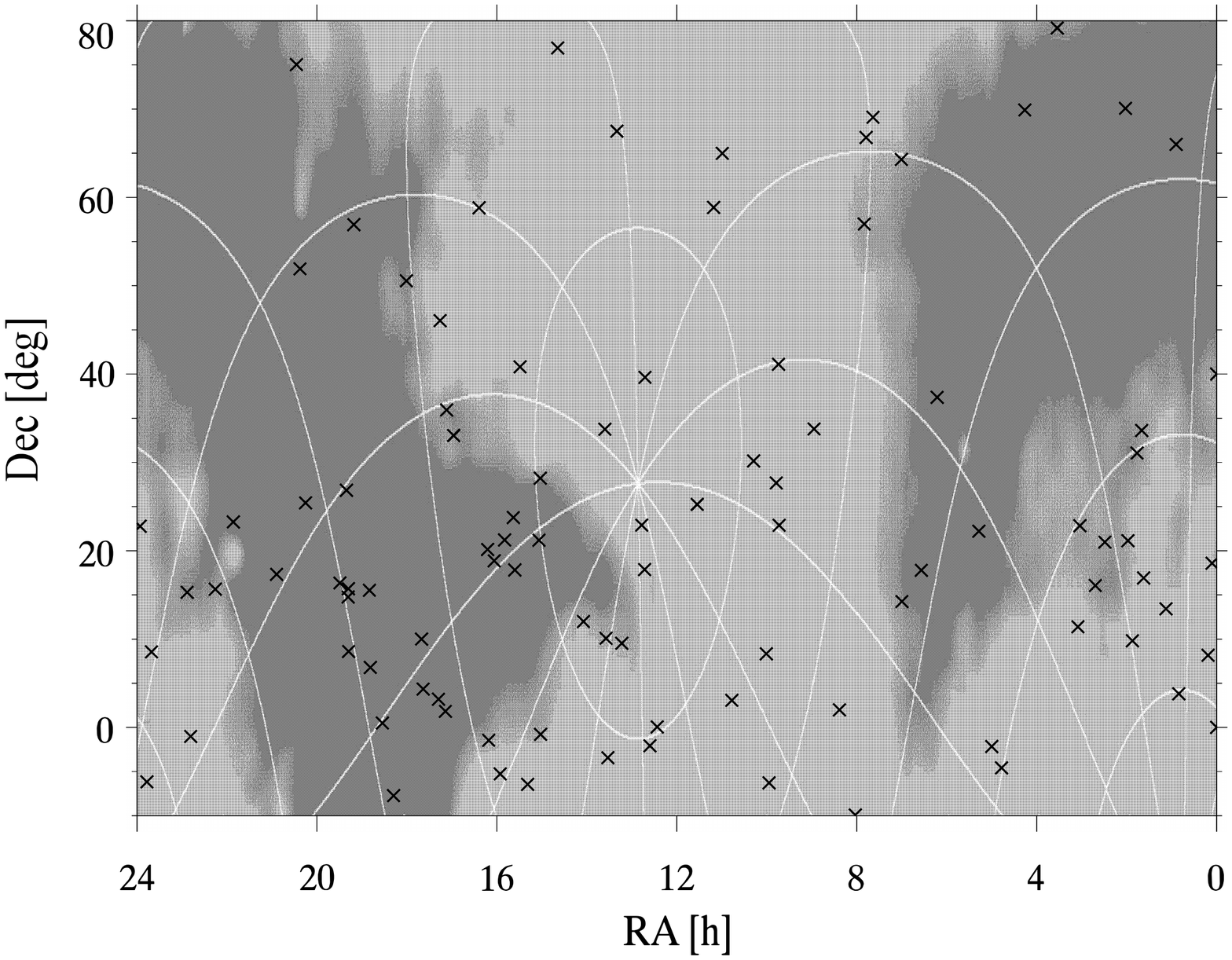} \\ b)}
    \end{minipage}
    \caption{A map of the celestial sphere with twenty pulsars detected in this work (black stars, a) and 102 not yet detected pulsars with the parameters selected for the census (black crosses, b). Positions of the pulsars are overlapped on the northern sky map at 20\,MHz \citep{Sidorchuk2008}}.
    \label{fig:map_detected.pdf}
\end{figure*}

\begin{figure}
	\includegraphics[width=\columnwidth]{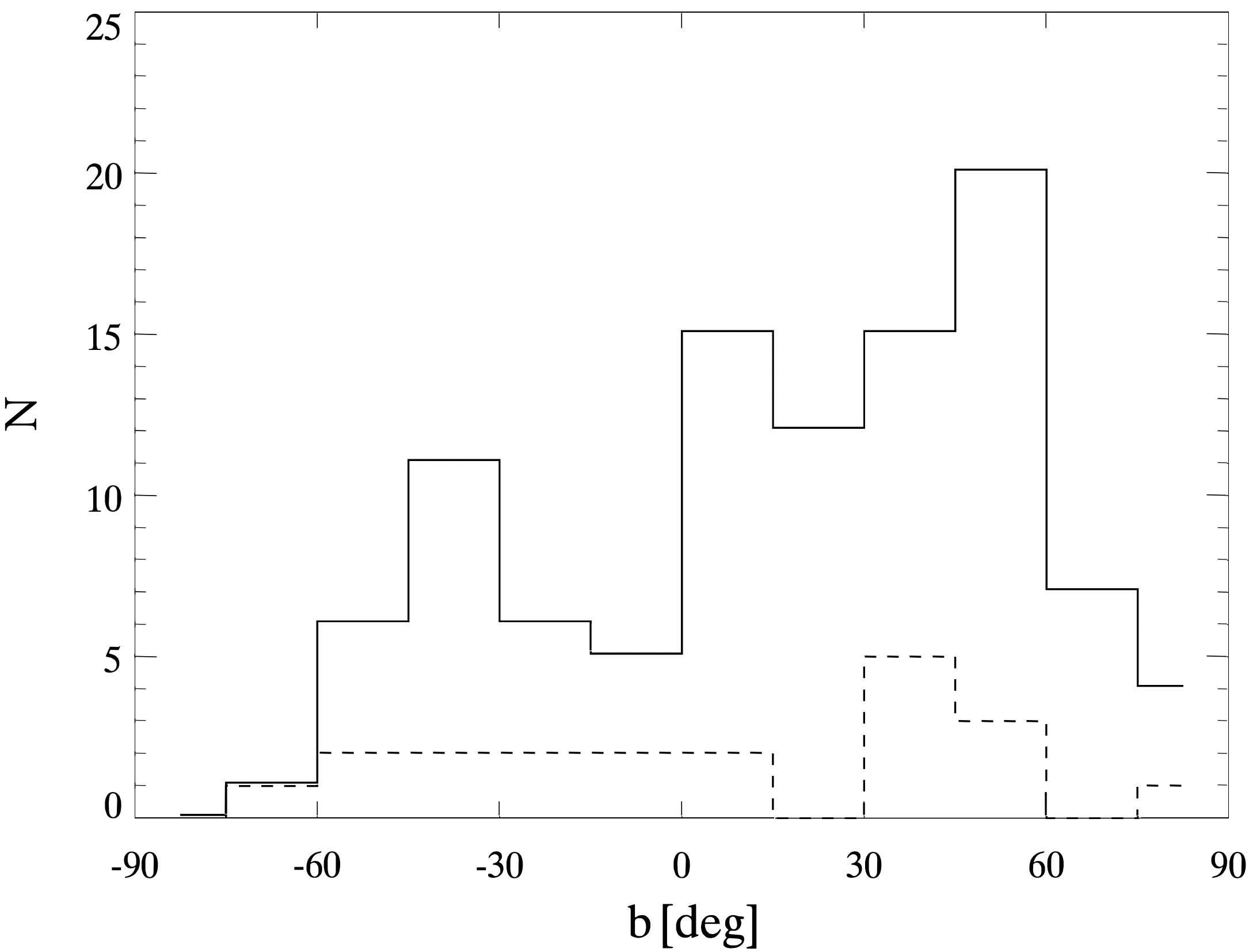}
    \caption{Galactic latitudes distribution histogram of detected (dashed line) and undetected pulsars (solid line). Each bin spans $15^{\circ}$ on x-axis range from $-90^{\circ}$ to $90^{\circ}$.}
    \label{Hist_det_nodet.pdf}
\end{figure}

In this study, we found decametre radio emission of 20 pulsars, thus, their total number in this wave range reached 63 for now. The left panel of Fig.~\ref{fig:map_detected.pdf} shows all 20 detected pulsars mapped to the celestial sphere, the right one – 102 not yet detected pulsars of the selected parameter range. Fig.~\ref{Hist_det_nodet.pdf} shows the histogram of galactic latitudes distribution of already detected and not yet detected pulsars. In decametre wave range, galactic background brightness temperature is very high, which significantly complicates the detection of pulsars. As one can see from Fig.~\ref{fig:map_detected.pdf}, about 2/3 of the 122 sought pulsars are located in rather cold regions of the sky. We obtained approximately the same ratio for the 20 newly detected sources. At the same time, this may indicate that the regions far from the Galactic plane have not been studied thoroughly enough. It means that we can expect new discoveries of nearby pulsars mainly in these regions.

At low frequencies, due to the wide beam of UTR-2 antenna cross-shaped pattern in the summation mode we use to search for pulsars, the antenna side lobes capture the "hot" regions of the sky. And one of the methods to reduce the noise level when observing near "hot" areas is using antenna pattern with non-rectangular current distribution between the UTR-2 sections, which reduces the side lobes, and which is the aim of our future work. 

Many pulsars we are tried to detect at decametre waves were discovered recently, so their timing needs improvement. As shown above, an inaccurate pulsar period is often an obstacle to its reliable detection in the low-frequency wave range. Therefore, we have improved our pulsar data processing and analysis routines and now they allow us to vary in a wide range both the pulsar rotation period and its dispersion measure. This processing pipeline will probably help us to discover the sources of repetitive radio emission (pulsars and RRATs) from the data of the first decametre survey of the Northern sky \citep{Vasylieva2013, Vasylieva2015,Kravtsov2016a,Kravtsov2016b, Kravtsov2016c, Zakharenko2018}.

\section{Conclusions}

The main result of the second decametre pulsar census is the discovery of low-frequency radio emission from 20 sources and the refinement of their radio emission parameters. The exact dispersion measure values will probably make it possible to identify some of the transient signals detected in the decametre wave range. A wide range of galactic latitudes, in which a significant number of the new pulsars were discovered, as well as rather high flux densities of some of them, give us a hope that further studies in the entire range of galactic latitudes at low frequencies will not be fruitless. This is also evidenced by the rapid (more than doubled in 10 years) increase in the number of newly discovered pulsars with a dispersion measure of up to $30$\,pc\,cm$^{-3}$. Therefore, planning and making new surveys in the entire galactic latitudes range, as well as increasing the sensitivity of such observations, is an urgent task especially at low frequencies. The ways for solving the last of the aforementioned problems are increasing of the observation time, improving algorithms of RFI mitigation and searching for both periodic and single cosmic pulsed signals. It will make possible to advance in understanding the population of neutron stars in the nearest galactic surroundings, as well as in understanding the processes of their formation and emission.

\section*{Acknowledgements}

This work is supported by a Grant for Research of Young Scientists of the NAS of Ukraine (2019-2020), a Grant of the NAS of Ukraine for Research Laboratories/Groups of Young Scientists of the NAS of Ukraine (2020-2021), as well as the Latvian State Fellowship for Research (2020).

\section*{Data Availability}

The data underlying this paper will be shared on reasonable request to the corresponding author.



\bibliographystyle{mnras}
\bibliography{library} 




\appendix


\bsp	
\label{lastpage}
\end{document}